\documentclass[epjCONF]{svjour}
\usepackage{graphicx}
\usepackage{bm}
\usepackage[varg]{txfonts} % Times fonts
\usepackage[latin1]{inputenc}
\session-title{%
19$^{\textnormal{\footnotesize th}}$ International %
IUPAP Conference on Few-Body Problems in Physics%
}

\newcommand{\bfr}{{\bm r}}

\newcommand{\bfk}{{\bm k}}

\newcommand{\bfp}{{\bm p}}

\newcommand{\bfx}{{\bm x}}
\newcommand{\bfy}{{\bm y}}
\newcommand {\bra}{\langle}
\newcommand {\ket}{\rangle}
\newcommand {\refeq}[1] {(\ref{#1})}

\begin{document}
\title{%
%Insert your title here: 
Recent Developments in Few-Nucleon Scattering
}%
\author{%
% add email of the responsible author:
A. Kievsky\thanks{\email{kievsky@pi.infn.it}} 
}
\institute{%
Istituto Nazionale di Fisica Nucleare, Largo B. Pontecorvo 3, 56127 (Pisa),
Italy }
\abstract{
Using modern nucleon-nucleon interactions in the description of the
$A=3,4$ nuclei, it is not possible to reproduce both the three-
and four-nucleon binding energies simultaneously. This is one manifestation of
the necessity of including a three-nucleon force in the nuclear Hamiltonian.
Several models of the three-nucleon force exist and are applied in the
description of light nuclei.
However, as it is discussed here, a simultaneous description of the
three- and four-body binding energies and the $n-d$ doublet scattering length 
seems to be problematic. Accordingly, a comparative study of some of these
models is performed.
In a different analysis, we study applications of the Kohn Variational
Principle, formulated in terms of
integral relations, to describe $N-d$ scattering processes.
} %end of abstract
\maketitle
%
% 
%----- Beginning of MAIN TEXT  --------------------------------------- 
% 
\section{Introduction}
\label{kievsky_intro}

%Your text comes here ...
Realistic nucleon-nucleon (NN) potentials reproduce the
experimental NN scattering data up to energies
of $350$ MeV with a $\chi^2$ per datum close to 1.
However, the use of these potentials in the description
of the three- and four-nucleon bound and scattering states gives a
$\chi^2$ per datum much larger than 1 (see for example 
Refs.\cite{walter,kiev01}).
In order to improve that situation, different three-nucleon
force (TNF) models have been introduced so far. Widely
used in the literature are the Tucson-Melbourne (TM) and the
Urbana IX (URIX) models \cite{tm,urbana}. These models are based
on the exchange mechanism of two pions between three nucleons with the
intermediate excitation of a $\Delta$ resonance.
The TM model has been revisited
within a chiral symmetry approach~\cite{friar:a},
and it has been demonstrated that the contact term present in it
should be dropped. This new TM potential, known as TM$'$, has been
subsequently readjusted~\cite{tmp}. The final operator structure
coincides with that one
given in the TNF of Brazil already derived many years
ago~\cite{brazil}.
Recently, TNFs have been derived based on chiral effective field theory at 
next-to-next-to-leading order~\cite{epelbaum02}.
A local version of these interactions (hereafter referred as N2LOL)
can be found in Ref.~\cite{N2LO}.
All these models contain a certain number of parameters that fix the
strength of the interaction. It is a common
practice to determine these parameters from
the three- and four-nucleon binding energies. A particular TNF is in
general associated to a specific NN potential and the sum of the two interactions
forms the nuclear potential energy. The two- and three-nucleon interactions 
derived using chiral effective field theory are consistently constructed.
However a particular TNF can be used associated with different NN interactions.
As a consequence, the parametrization of a particular TNF could change 
since different NN potentials predict different $A=3,4$ binding energies.

More recently, a new class of two-nucleon interactions has been
obtained ($V_{low-k}$ potentials). With the purpose of
eliminating the high-momentum part of the interaction, 
the Hilbert space has been separated
into low and high momentum regions and the renormalization group
method has been used to integrate out the high momentum components
above a cutoff $\Lambda$~\cite{Bog07}.
The value for $\Lambda$ is typically chosen to reproduce
the triton binding energy.

All these potential models can be used to study bound and 
scattering states in the $A=3,4$ systems in order to extract
information about their capability to describe the nuclear
dynamics. Besides the bound state energies,
in the $A=3$ system,
the $n-d$ doublet scattering length $^2a_{nd}$ can give valuable information.
In principle this quantity is correlated, to some extent, to the $A=3$ binding
energy through the so-called Phillips line~\cite{phillips,bedaque}.
However the presence of TNFs of the type studied here
breaks this correlation. Therefore $^2a_{nd}$ emerges as an independent
observable. Due to the lack of excited states in the
$A=3$ system, the zero energy state is the first one above the ground state.
In the case of $n-d$ scattering at zero energy, the $J={\frac{1}{2}}^+$ state
is orthogonal to the
triton ground state and, for this reason, it presents a node in the relative
distance between the incident nucleon and the deuteron. The position of the
node is related to the scattering length and it is also
sensitive to the relation between the overall attraction and repulsion of the
interaction. Several of the realistic NN potentials underestimate
the triton binding energy. Therefore by adding a TNF,
with the strength fixed for example to reproduce the triton binding energy,
the balance between the total attraction and repulsion in the potential
changes. This leads to a modification in $^2a_{nd}$ and
this modification depends on the parameters in the TNF.
The determination
of the TNF parametrization able to describe the triton binding energy
$B$($^3$H), the $\alpha$-particle binding energy $B$($^4$He) and $^2a_{nd}$ has
been analyzed in Ref.~\cite{epelbaum02} for a TNF derived from chiral effective
field theory. A similar analysis has not been done for the local TNF models 
URIX, TM' and N2LOL.

In Refs.~\cite{report,marcucci09} results for $B$($^3$H), $B$($^4$He), $^2a_{nd}$
are given using different combinations of NN interactions
(see Table~\ref{tb:table1}). Those results indicate
that the models are not able to describe simultaneously the $A=3,4$
binding energies and $^2a_{nd}$. 
In order to analyze further the mentioned discrepancies, here we study
potential models constructed summing to the AV18 NN potential~\cite{av18} 
the three-nucleon interactions of TM', URIX and N2LOL. 
Parametrizations of the URIX and TM'
models already exist in conjunction with the AV18 potential. Conversely the
N2LOL force has been constructed using the N3LO-Idaho potential 
from Entem et al.~\cite{entem}. 
So, here we adapt its parametrization to reproduce,
when summed to the AV18 interaction, the triton binding energy.
Different parametrizations of the three TNF models are analyzed studying
the description of $B$($^3$H), $B$($^4$He) and $^2a_{nd}$ and 
some polarization observables in $p-d$ scattering.
The calculations have been done using the hyperspherical
harmonic (HH) method as given in Refs.~\cite{phh,kiev97,hh4b,hh4s}
to describe bound and scattering states in $A=3,4$ systems using local
potentials. The extension
to treat nonlocal potentials was given in Refs.~\cite{marcucci09,viv06}.

In a different application devoted to study scattering states
in few-nucleon systems, a discussion of the use of the integral
relations derived in Ref.~\cite{intrel} from the Kohn Variational
principle (KVP) is given. 
It has been shown that starting from the KVP, the tangent of the phase-shift
can be put in a form of a quotient where both, the numerator and the denominator,
are given in the form of an integral relation. This is similar to what was proposed
in Ref.~\cite{harris}, however its strict relation with the KVP has not been 
recognized. To be noticed that a
general formulation of the scattering theory using surface-integrals is given
in Ref.~\cite{kadyrov}.
Here we would like to discuss some specific examples of the integral relations
derived from the KVP. Starting the analysis in the simplest case,
the $A=2$ system, we show that they can be used to compute phase-shifts from
bound state like functions. A second application of the integral
relations regards the possibility of determining
phase-shifts from a calculation in which the Coulomb
potential has been screened. All these examples serve
to demonstrate the general validity of the KVP formulated in terms
of integral relations. Due to their short-range nature, they are determined
by the wave function in the interaction region and not from its
explicit asymptotic behaviour. This means that each wave function
solving $(H-E)\Psi=0$ in the interaction region can be used to determine
the corresponding scattering amplitude even if its asymptotic behaviour is not
the physical one.

\section{The HH expansion for $A=3,4$ systems}
\label{sec:form}

In this section we briefly review the HH method for bound and
scattering states. 

\subsection{The HH Method for Bound States}
\label{subsec:bs}

The nuclear wave function for the three-body system can be written as 
\begin{equation}
|\Psi\rangle=\sum_\mu c_\mu |\Psi_\mu\rangle \ ,
\label{eq:psi}
\end{equation}
where $|\Psi_\mu\rangle$ is a suitable complete set of 
states, and $\mu$ is an index denoting the set of quantum numbers 
necessary to completely specify the basis elements. 

The coefficients of the expansion can be calculated using the 
Rayleigh-Ritz variational principle, which states that
\begin{equation}
  \langle\delta_c \Psi\,|\,H-E\,|\Psi\rangle
   =0 \ ,
   \label{eq:rrvar}
\end{equation}
where $\delta_c \Psi$ indicates the variation of 
$\Psi$ for arbitrary infinitesimal 
changes of the linear coefficients $c_\mu$. Where
the Hamiltonian of the system consists in the kinetic part plus
two- and three-nucleon interaction terms
\begin{equation}
  H=-\frac{\hbar^2}{2m}\sum_i\nabla^2_i +\sum_{i<j}V(i,j)+\sum_{i<j<k}W(i,j,k) 
\end{equation}

The problem of determining $c_\mu$ and the energy $E$ 
is reduced to a generalized eigenvalue problem, 
\begin{equation}
  \sum_{ \mu'}\,\langle\Psi_\mu\,|\,H-E\,|\, \Psi_{\mu'}\,\rangle \,c_{\mu'}=0
  \ .
  \label{eq:gepb}
\end{equation}
The main difficulty of the method is to compute the 
matrix elements of the Hamiltonian $H$ with respect to the basis states
$|\Psi_\mu\rangle$. Usually $H$ is given as a sum of terms (kinetic energy,
two-body potential, etc.). The calculation of the matrix elements of
some parts of $H$ can be more conveniently performed in coordinate 
space, while for other parts it could be easier to work in momentum
space. Therefore, it is important that the basis states 
$|\Psi_\mu\rangle$ have simple expressions in both spaces. The
HH functions indeed have such a property.

In the case of three nucleons of mass
$m$ the Jacobi vectors $\bfx_{1p},\bfx_{2p}$
correspond to a given particle permutation denoted with $p$, which 
specifies the particle order 
$i,j,k$, 
\begin{eqnarray}
  \bfx_{2p}&=&\frac{1}{\sqrt{2}}(\bfr_j-\bfr_i) \ , \nonumber \\
  \bfx_{1p}&=&\sqrt{\frac{2}{3}}(\bfr_k-\frac{1}{2}(\bfr_i+\bfr_j)) \ . 
  \label{eq:jacc3}
\end{eqnarray}
Here $p=1$ corresponds to the order 1,2,3.
It is convenient to replace the modulii of $\bfx_{2p}$ and 
$\bfx_{1p}$ with the so-called hyperradius and 
hyperangle, defined as
\begin{eqnarray}
  \rho&=&\sqrt{\bfx_{1p}^2+\bfx_{2p}^2} \ , \label{eq:rho} \\
  \tan{\phi_{p}}&=&\frac{x_{1p}}{x_{2p}} \ .  \label{eq:hypera}
\end{eqnarray}
Note that $\rho$ does not depend on
the particle permutation $p$. 
The complete set of hyperspherical coordinates is then 
given by $\{\rho,\Omega^{(\rho)}_p\}$, with 
\begin{equation} 
  \Omega^{(\rho)}_p=[{\hat{\bfx}}_{1p},{\hat{\bfx}}_{2p};\phi_{p}] \ , 
  \label{eq:omegar}
\end{equation}
and the suffix $(\rho)$ recalls the use of the coordinate space.  

The expansion states $|\Psi_\mu\rangle$ of 
Eq.~(\ref{eq:psi}) are then given by 
\begin{equation}
  |\,\Psi_\mu^{(\rho)}\,\rangle
  = f_l(\rho) {\cal Y}_{ \{G\} }(\Omega^{(\rho)})\ ,
  \label{eq:rexp}
\end{equation}
where $f_l(\rho)$ for $l=1,\ldots\,M$ is a complete set of hyperradial 
functions, chosen of the form
\begin{equation}
 f_l(\rho)=\gamma^{3} \sqrt{\frac{l!}{(l+5)!}}\,\,\, 
 L^{(5)}_l(\gamma\rho)\,\,{\rm e}^{-\frac{\gamma}{2}\rho} \ .
 \label{eq:fllag}
\end{equation}
Here $L^{(5)}_l(\gamma\rho)$ are Laguerre polynomials, 
and the non-linear parameter $\gamma$ is 
variationally optimized. As an example, for the N3LO-Idaho potential, 
it can be chosen in the interval 6--8 fm$^{-1}$.

The functions ${\cal Y}_{ \{G\} }(\Omega^{(\rho)})$ are written 
as 
\begin{equation}
  {\cal Y}_{ \{G\} }(\Omega^{(\rho)})= \sum_{p=1}^{3}\bigg[ 
  Y^{LL_z}_{ [G] }(\Omega^{(\rho)}_p) 
  \otimes [S_2\otimes \frac{1}{2}]_{S S_z} \bigg]_{J J_z}\, 
  [T_2\otimes\frac{1}{2}]_{T T_z}   \ , \label{eq:hha3}
\end{equation}
where the sum is performed over the three even permutations.  
The spin (isospin) of particles $i$ and $j$ are coupled to 
$S_2$ ($T_2$), which is itself coupled to the spin (isospin) 
of the third particle to give 
the state with total spin $S$ (isospin $T,T_z$).
The total orbital angular momentum $L$ and the total 
spin $S$ are coupled to the total angular momentum 
$J,J_z$.  
The functions $Y^{LL_z}_{[G]}(\Omega^{(\rho)}_p)$, 
having  a definite value of 
$L,L_z$, are the HH functions:
\begin{equation}
  Y^{LL_z}_{ [G] }(\Omega^{(\rho)}_p) =
  \biggl[Y_{\ell_2}({\hat{\bfx}}_{2p}) \otimes
   Y_{\ell_1}({\hat{\bfx}}_{1p}) \biggr]_{LL_z}
  N_{[G] }\, ^{(2)}P_{n}^{\ell_1,\ell_2}(\phi_p) \ .
\label{eq:hh3}
\end{equation}
Here $Y_{\ell_1}({\hat{\bfx}}_{1p})$ and $Y_{\ell_2}({\hat{\bfx}}_{2p})$ are 
spherical harmonics, $N_{[G]}$ is a normalization factor and 
$^{(2)}P_{n}^{\ell_1,\ell_2}(\phi_p)$ is an hyperspherical polynomial.
The grand angular quantum number $G$ is 
defined as $G=2n+\ell_1+\ell_2$.
The notations $[G]$ and $\{G\}$ of Eqs.~(\ref{eq:hh3}) 
and~(\ref{eq:hha3}) stand for $[\ell_1,\ell_2;n]$ and 
$\{\ell_1,\ell_2,L,S_2,T_2$, $S,T;n\}$, respectively, 
and $\mu$ of Eq.~(\ref{eq:rexp}) is $\mu=\{G\},l$. Note that 
each set of quantum numbers $\{\ell_1,\ell_2,L,S_2,T_2,S,T\}$ 
is called ``channel'', and the antisymmetrization of 
${\cal Y}_{ \{G\} }(\Omega^{(\rho)})$ requires $\ell_2+S_2+T_2$ 
to be odd. In addition, $\ell_1+\ell_2$ must be even (odd) 
for positive (negative) parity.

The HH functions having grand angular quantum number $G$ constructed in
terms of a given set of Jacobi vectors
$\bfx_{1p},\bfx_{2p}$, defined starting from the particle 
order $i,j,k$, can always be expressed in terms of
the HH functions constructed, for instance, in terms of
$\bfx_{1 (p=1)},\bfx_{2 (p=1)}$ with the same value of $G$.  
In fact, the following relation holds
\begin{equation}
  Y^{LL_z}_{[\ell_1,\ell_2;n]}(\Omega^{(\rho)}_p) =
   \sum_{\ell_1',\ell_2',n'}
    a^{(p),L}_{\ell_1,\ell_2,n;\,\ell_1',\ell_2',n'}
    Y^{LL_z}_{[\ell_1',\ell_2';n']}(\Omega^{(\rho)}_{(p=1)})\ ,
   \label{eq:rr3}
\end{equation}
where the sum is
restricted to the values $\ell_1'$, $\ell_2'$, and $n'$
such that $\ell_1'+\ell_2'+2n'=G$. 
The coefficients $ a^{(p),L}_{\ell_1,\ell_2,n;\,\ell_1',\ell_2',n'}$
relating the two sets of HH functions are known as the Raynal-Revai
coefficients~\cite{RR70}.
Also the spin-isospin states can be recoupled to obtain states where the
spin and isospin quantum numbers are coupled in a given order of the particles. 
The result is that the antisymmetric functions 
${\cal Y}_{ \{G\} }$ can be expressed as a
superposition of functions constructed in terms of a given order of particles
$i,j,k$, each one having the pair $i$,$j$ in a definite spin and
angular momentum state. When the two-body potential acts on the pair of
particles $i$,$j$, the effect of the projection is easily taken into
account. 

The expansion states of Eq.~(\ref{eq:psi}) 
in momentum space can be obtained as follows.
Let $\hbar\bfk_{1p},\hbar\bfk_{2p}$ be the conjugate Jacobi momenta 
of the Jacobi vectors, given by 
\begin{eqnarray}
  \hbar\bfk_{2p}&=&\frac{1}{\sqrt{2}}(\bfp_j-\bfp_i) \ , \nonumber \\
  \hbar\bfk_{1p}&=&\sqrt{\frac{2}{3}}(\bfp_k-\frac{1}{2}(\bfp_i+\bfp_j)) \ , 
  \label{eq:jacm3}
\end{eqnarray}
$\bfp_i$ being the momentum of the $i$-th particle. 
We then define a hypermomentum $Q$ and a set of 
angular-hyperangular variables as 
\begin{eqnarray}
  Q&=&\sqrt{\bfk_{1p}^2+\bfk_{2p}^2} \ ,   \nonumber \\
  \Omega^{(Q)}_p&=&[{\hat{\bfk}}_{2p},{\hat{\bfk}}_{1p};\varphi_{p}] \ , 
  \label{eq:hyperq}
\end{eqnarray}
where
\begin{equation}
  \tan{\varphi_{p}}=\frac{k_{1p}}{k_{2p}} \ .
  \label{eq:hyperaq}
\end{equation}
Then, the momentum-space version of the wave function 
given in Eq.~(\ref{eq:rexp}) is
\begin{equation}
  |\,\Psi_\mu^{(Q)}\,\rangle=
   g_{ G,l }(Q) {\cal Y}_{ \{G\} }(\Omega^{(Q)}) \ , 
  \label{eq:qexp}
\end{equation}
where ${\cal Y}_{ \{G\} }(\Omega^{(Q)})$ is the same as 
${\cal Y}_{ \{G\} }(\Omega^{(\rho)})$ of Eq.~(\ref{eq:hha3})
with $\bfx_{ip}\rightarrow\bfk_{ip}$, 
and 
\begin{equation}
   g_{G,l}(Q)=(-i)^G\,\int_0^\infty d\rho\,
   \frac{\rho^{3}}{Q^{2}}\,
   J_{G+2}(Q\rho)\, f_{l}(\rho) \ .
\label{eq:vg}
\end{equation}
With the adopted form of $f_l(\rho)$ given in Eq.~(\ref{eq:fllag}), 
the corresponding functions $g_{G,l}(Q)$ can be easily calculated, 
and they are explicitly given in Ref.~\cite{viv06}.

\subsection{The HH Method for Scattering States Below Deuteron 
Breakup Threshold}
\label{subsec:ss}

We consider here the extension of the 
HH technique to describe $N-d$ scattering states 
below deuteron breakup threshold, when both local 
and non-local interaction models are considered.

The wave function $\Psi_{N-d}^{L S J J_z}$ describing the $N-d$
scattering state with incoming orbital angular momentum $L$ and channel spin
$S$, parity $\pi=(-)^L$, 
and total angular momentum $J, J_z$,
can be written as 
\begin{equation}
    \Psi_{N-d}^{LSJJ_z}=\Psi_C^{LSJJ_z}+\Psi_A^{LSJJ_z} \ ,
    \label{eq:psica}
\end{equation}
where $\Psi_C^{LSJJ_z}$ describes the system in the region where the particles
are close to each other and their mutual interactions are strong, 
while $\Psi_A^{LSJJ_z}$ describes the relative motion between the nucleon $N$
and the deuteron in the asymptotic region, where the $N-d$ nuclear 
interaction is negligible. The function $\Psi_C^{LSJJ_z}$, which has to 
vanish in the limit of large intercluster
separations, can be expanded on the HH basis as it has been done 
in the case of bound states. Therefore, 
applying Eq.~(\ref{eq:psi}), 
the function $\Psi_C^{LSJJ_z}$ can be casted in the form 
\begin{equation}
  |\Psi^{LSJJ_z}_C\rangle=\sum_{\mu}\, c_\mu\,
  |\Psi_\mu \rangle \ ,
  \label{eq:psis}
\end{equation}
where $|\Psi_\mu\rangle$ is defined in Eqs.~(\ref{eq:rexp}) 
and~(\ref{eq:qexp}) in coordinate- and momentum-space, respectively.

The function $\Psi_A^{LSJJ_z}$ is the appropriate 
asymptotic solution of the relative $N-d$ Schr\"odinger equation. 
It is written as a linear combination of the following functions, 
\begin{equation}
  \Omega_{LSJJ_z}^{\lambda}=\sum_{p=1}^3\Omega_{LSJJ_z}^{\lambda}(p) \ ,
  \label{eq:psiomp} 
\end{equation}
where the sum over $p$ has to be done over the three even permutations 
and
\begin{eqnarray}
  \Omega_{LSJJ_z}^{\lambda}(p)&=& \sum_{l=0,2}w_l(x_{2p})\,R^\lambda_L(y_p)
\Bigl\{\Bigl[ [Y_l(\hat{\bfx}_{2p})\otimes S_2]_1\otimes \frac{1}{2}\Bigr]_S
\nonumber \\ & & \otimes Y_{L}(\hat{\bfy}_p) \Bigr \}_{JJ_z} 
[ T_2\otimes \frac{1}{2} ]_{TT_z} \ .
%\Bigl [ [\chi_N\otimes \phi_d]_{S} \otimes 
%  Y_{L}(\hat{y}_p) \Bigr ]_{JJ_z} {\cal R}^\lambda_L(y_p) \ ,
  \label{eq:psiom}
\end{eqnarray}
Here the spin and isospin quantum numbers of particles $i$ and $j$ 
have been coupled to $S_2$ and $T_2$, with $S_2=1$, $T_2=0$ for the 
deuteron,  
$w_l(x_{2p})$ is the deuteron wave function component in the waves $l=0,2$, 
${\bfy}_p$ is the distance 
between $N$ and the center of mass of the deuteron, i.e. 
$\bfy_p=\sqrt{\frac{3}{2}}\bfx_{1p}$, 
$Y_l(\hat{\bfx}_{2p})$ and $Y_{L}(\hat{\bfy}_p)$ are the standard spherical 
harmonic functions, 
and the functions $R^\lambda_L(y_p)$ are the regular ($\lambda\equiv R$)
and irregular ($\lambda\equiv I$) radial solutions of the relative two-body 
$N-d$ Schr\"odinger equation without the nuclear interaction. 
These regular and irregular functions, denoted as 
${\cal F}_L(y_p)$ and ${\cal G}_L(y_p)$ respectively, have the form
\begin{eqnarray}
{\cal F}_L(y_p)&=&\frac{1}{(2L+1)!!q^L C_L(\eta)}\,{F_L(\eta,\xi_p)\over \xi_p} \ , 
\nonumber \\
{\cal G}_L(y_p)&=&(2L+1)!! q^{L+1}C_L(\eta)f_R(y_p){G_L(\eta,\xi_p)\over \xi_p}
\ ,
\label{eq:risol}
\end{eqnarray}
where $q$ is the
modulus of the $N-d$ relative momentum 
(related to the total kinetic energy in the center of mass system by
$T_{cm}={q^2\over 2\mu}$, $\mu$ being the $N-d$ reduced mass), 
$\eta=2\mu e^2/q$ and $\xi_p=qy_p$ are the usual Coulomb parameters, 
and the regular (irregular) Coulomb function $F_L(\eta,\xi_p)$ 
($G_L(\eta,\xi_p)$) and the 
factor $C_L(\eta)$ are defined in the standard 
way~\cite{chen:b}. The factor $(2L+1)!! q^L C_L(\eta)$
has been introduced so that ${\cal F}$ and ${\cal G}$
have a finite limit for $q\rightarrow 0$.
The function $f_R(y_p)=[1-\exp(-b y_p)]^{2L+1}$ 
has been introduced to regularize $G_L$ at small values of $y_p$. 
The trial parameter $b$ is
determined by requiring that $f_R(y_p)\rightarrow 1$ outside
the range of the nuclear interaction,
thus not modifying the asymptotic behaviour of the 
scattering wave function. A value of $b=0.25$ fm$^{-1}$ 
has been found appropriate.
The non-Coulomb case of Eq.~(\ref{eq:risol}) is
obtained in the limit $e^2\rightarrow 0$. In this case, $F_L(\eta,\xi_p)/\xi_p$ and
$G_L(\eta,\xi_p)/\xi_p$ reduce to the regular and irregular Riccati-Bessel 
functions and
the factor $(2L+1)!!C_L(\eta)\rightarrow 1$ for $\eta\rightarrow 0$.

With the above definitions, $\Psi_A^{LSJJ_z}$  can be written in the form
\begin{equation}
  \Psi_A^{LSJJ_z}= \sum_{L^\prime S^\prime}
 \bigg[\delta_{L L^\prime} \delta_{S S^\prime} 
\Omega_{L^\prime S^\prime JJ_z}^R
  + {\cal R}^J_{LS,L^\prime S^\prime}(q)
     \Omega_{L^\prime S^\prime JJ_z}^I \bigg] \ ,
  \label{eq:psia}
\end{equation}
where the parameters ${\cal R}^J_{LS,L^\prime S^\prime}(q)$ give the
relative weight between the regular and irregular components 
of the wave function. They
are closely related to the reactance matrix (${\cal K}$-matrix)
elements, which can be written as
\begin{eqnarray}
 & {\cal K}^J_{LS,L^\prime S^\prime}(q)= &\cr
 & (2L+1)!!(2L'+1)!!&q^{L+L'+1}C_L(\eta)C_{L^\prime}(\eta)
 {\cal R}^J_{LS,L^\prime S^\prime}(q) \;\;\ .
\end{eqnarray}
By definition of the ${\cal K}$-matrix, its eigenvalues are
$\tan\delta_{LSJ}$, $\delta_{LSJ}$ being the phase shifts.
The sum over $L^\prime$ and $S^\prime$ in Eq.~(\ref{eq:psia}) is over all 
values compatible with a given $J$ and parity $\pi$. In particular, the sum 
over $L^\prime$ is limited to include either even or odd values since
$(-1)^{L^\prime}=\pi$.

The matrix elements ${\cal R}^J_{LS,L^\prime S^\prime}(q)$ and 
the linear coefficients $c_\mu$ occurring in the expansion of $\Psi^{LSJJ_z}_C$ 
of Eq.~(\ref{eq:psis})
are determined applying the Kohn variational principle, 
which states that the functional
\begin{eqnarray}
   [{\cal R}^J_{LS,L^\prime S^\prime}(q)]&=&
    {\cal R}^J_{LS,L^\prime S^\prime}(q)
     - \left \langle \Psi^{L^\prime S^\prime JJ_z }_{N-d} \left |
         {\cal L} \right |
        \Psi^{LSJJ_z}_{N-d}\right \rangle \ , \nonumber \\
{\cal L}&=&\frac{m}{2\sqrt{3}\hbar^2}(H-E) \ , \label{eq:kohn}
\end{eqnarray}
has to be stationary with respect to variations of the trial parameters 
in $\Psi^{LSJJ_z}_{N-d}$. 
Here $E$ is the total energy of the system, $m$ is the nucleon mass, 
and ${\cal L}$ is chosen so that 
\begin{equation}
   \langle \Omega^R_{LSJJ_z}| {\cal L} | \Omega^I_{LSJJ_z} \rangle
  -\langle \Omega^I_{LSJJ_z}| {\cal L} | \Omega^R_{LSJJ_z} \rangle =1 \ .
\end{equation}
As described in Ref.~\cite{kiev97}, 
using Eqs.~(\ref{eq:psis}) and~(\ref{eq:psia}), 
the variation of the diagonal functionals of Eq.~(\ref{eq:kohn}) with
respect to the linear parameters $c_\mu$ leads to the following 
system of linear inhomogeneous equations:
\begin{equation}
  \sum_{\mu'} \langle \Psi_\mu| {\cal L} |\Psi_{\mu'}\rangle c_{\mu'} = 
     -D^\lambda_{LSJJ_z}(\mu) \ .
  \label{eq:set1}
\end{equation}
Two different terms $D^\lambda$ corresponding to
$\lambda\equiv R,I$ are introduced and are defined as 
\begin{equation}
  D^\lambda_{LSJJ_z}(\mu)= \langle \Psi_\mu| {\cal L} |
\Omega^\lambda_{LSJJ_z}\rangle \ .
\label{eq:dlm}
\end{equation}
The matrix elements ${\cal R}^J_{LS,L'S'}(q)$ are obtained 
varying the diagonal functionals of Eq.~(\ref{eq:kohn}) with respect to them. 
This leads to the following set of algebraic equations
\begin{equation}
  \sum_{L'' S''} {\cal R}^J_{LS,L''S''}(q) X_{L'S',L''S''}= Y_{LS,L'S'} \ ,
\label{eq:set2}
\end{equation}
with the coefficients $X$ and $Y$ defined as
\begin{eqnarray}
X_{LS,L'S'}&= \langle
\Omega^I_{LSJJ_z}+\Psi^{LSJJ_z,I}_C| {\cal L} |\Omega^I_{L'S'JJ_z}\rangle \ ,
\nonumber \\
Y_{LS,L'S'}&=-\langle
\Omega^R_{LSJJ_z}+\Psi^{LSJJ_z,R}_C| {\cal L} |\Omega^I_{L'S'JJ_z}\rangle \ .
\label{eq:xy}
\end{eqnarray}
Here $\Psi^{LSJJ_z,\lambda}_C$ is the solution of the set of 
Eq.~(\ref{eq:set1}) with the corresponding $D^\lambda$ term. A 
second order estimate of ${\cal R}^J_{LS,L'S'}(q)$ is 
given by the quantities $[{\cal R}^J_{LS,L'S'}(q)]$, obtained by 
substituting in Eq.~(\ref{eq:kohn}) the
first order results. Such second-order calculation provides a symmetric 
reactance matrix. This condition is not {\it a priori} imposed, 
and therefore it is a useful test of the numerical accuracy.

In the particular case of $q=0$ (zero-energy scattering),
the scattering can occur only in the channel $L=0$ and the observables 
of interest are the scattering lengths. Within the 
present approach, they can be easily obtained from the relation
\begin{equation}
  ^{(2J+1)}a_{Nd}=-\lim_{q\rightarrow 0}{\cal R}^J_{0J,0J}(q)\ .
\label{eq:scleng}
\end{equation}

An alternative way to solve the scattering problem, used when 
$q\neq 0$, is to apply the complex Kohn variational principle 
to the ${\cal S}$-matrix, as in Ref.~\cite{kiev97}. 

The approach presented so far for bound and scattering states
does not have too many differences compared to the method 
presented for instance in Ref.~\cite{phh}, and known as 
pair-correlated hyperspherical harmonics (PHH) method. In fact, 
in the PHH method a correlation factor is included in the HH 
expansion of Eq.~(\ref{eq:psis}) to take into account the strong 
short-range correlations induced by the realistic two-body potentials, 
like the AV18. The presence of correlation 
functions makes the convergence of the expansion much faster than in the 
uncorrelated case. However, the PHH method cannot be simply 
implemented when non-local two-body interactions are considered, 
unless the Fourier transform of the potential is 
performed. The calculation involving $\Psi_C^{LSJJ_z}$
can be performed with the HH or PHH expansions 
in coordinate- or in momentum-space, depending 
on what is more convenient.

\section{Three Nucleon Force Models}

\label{kievsky_sec:1}
In Ref.~\cite{report} the description of bound states and zero-energy
states for $A=3,4$ has been reviewed in the context of the HH method.
In Table~\ref{tb:table1} we report results for
the triton and $^4$He binding energies as well as for the doublet
$n-d$ scattering length $^2a_{nd}$ using the
AV18 and the N3LO-Idaho NN potentials and using the following
combinations of two- and three-nucleon interactions:
AV18+URIX, AV18+TM', N3LO-Idaho+N2LOL and N3LO-Idaho+URIXp. In this last
model the parameter in front of the spin-isospin independent part of the 
URIX potential has been rescaled by a factor of 0.384 to fit the triton binding
energy~\cite{marcucci09} (we call this model URIXp). We have considered also
the $V_{low k}$ model, obtained from the AV18 interaction with a cutoff
parameter $\Lambda=2.2$ fm$^{-1}$.
The results are compared to the experimental values reported
in the table. Worthy of notice is
the recent very accurate datum for $^2a_{nd}$~\cite{doublet}.

\begin{table}[h]
\caption{The triton and $^4$He binding energies $B$ (in MeV),
and  doublet scattering length $^2a_{nd}$ (in fm)
calculated using the indicated two- and three-nucleon interactions.
The experimental results are also reported.}
\label{tb:table1}
\begin{tabular}{@{}llll}
\hline
Potential & $B$($^3$H) & $B$($^4$He) & $^2a_{nd}$ \cr
\hline
AV18             & 7.624    & 24.22   & 1.258 \cr
N3LO-Idaho       & 7.854    & 25.38   & 1.100 \cr
AV18+TM'         & 8.440    & 28.31   & 0.623 \cr
AV18+URIX        & 8.479    & 28.48   & 0.578 \cr
N3LO-Idaho+N2LOL & 8.474    & 28.37   & 0.675 \cr
N3LO-Idaho+URIXp & 8.481    & 28.53   & 0.623 \cr
$V_{low-k}$      & 8.477    & 29.15   & 0.572 \cr
\hline
Exp.             & 8.482    & 28.30   & 0.645$\pm$0.003$\pm$0.007 \cr
\hline
\end{tabular}
\end{table}

From the table we may observe that only the results obtained
using an interaction model that includes a TNF are close to the
corresponding experimental values. In the case of the AV18+TM', the
strength of the TM' potential has been fixed to reproduce the
$^4$He binding energy and, as can be seen from the table, the
triton binding energy is underpredicted. Conversely,
the strength of the URIX potential
has been fixed to reproduce the triton binding energy giving too much
binding for $^4$He. The strength of the N2LOL potential has been
fixed to reproduce simultaneously the triton and the $^4$He binding
energies whereas the N3LO-Idaho+URIXp model overbinds $^4$He. These two
models give a better description of $^2a_{nd}$. The $V_{low-k}$ interaction
reproduces the triton binding energy but overbinds $^4$He appreciably
and $^2a_{nd}$ is not well described. In conclusion a simultaneous
correct description of the three quantities is not achieved by any of
the models considered.

To analyze further this fact,
we give a brief description of the TM' (or Brazil),
URIX and N2LOL models.  They can be put in the following way:
\begin{eqnarray}
W(1,2,3) & = & aW_a(1,2,3)+bW_b(1,2,3)+dW_d(1,2,3) \nonumber \\
W(1,2,3) & = & aW_a(1,2,3)+bW_b(1,2,3)+dW_d(1,2,3) \nonumber \\
& & +c_DW_D(1,2,3)+c_EW_E(1,2,3) \; .
\label{eq:w123}
\end{eqnarray}
Each term corresponds to a different source and has a different operator
structure.
The first three terms arise from the exchange of two pions between three nucleons.
The $a$-term is coming from $\pi N$ $S$-wave scattering
whereas the $b$-term and $d$-term, which are the most important,
come from $\pi N$ $P$-wave scattering. The specific form of these three terms
in configuration space is the following:
\begin{eqnarray}
 W_a(1,2,3) & = & \frac{W_0}{c^2\hbar^2}
 (\bm\tau_1\cdot\bm\tau_2)(\bm\sigma_1\cdot \bm r_{31})
 (\bm\sigma_2\cdot \bm r_{23}) y(r_{31})y(r_{23}) \nonumber \\
W_b(1,2,3) & = & W_0 (\bm\tau_1\cdot\bm\tau_2) [(\bm\sigma_1\cdot\bm\sigma_2)
  y(r_{31})y(r_{23})  \nonumber \\
 &+ & (\bm\sigma_1\cdot \bm r_{31})
    (\bm\sigma_2\cdot \bm r_{23})(\bm r_{31}\cdot \bm r_{23})
  t(r_{31})t(r_{23}) \nonumber \\
 &+ & (\bm\sigma_1\cdot \bm r_{31})(\bm\sigma_2\cdot \bm r_{31})
  t(r_{31})y(r_{23}) \nonumber \\
 &+ & (\bm\sigma_1\cdot \bm r_{32})(\bm\sigma_2\cdot \bm r_{32})
  y(r_{31})t(r_{23})]  \\
 W_d(1,2,3) & = & W_0(\bm\tau_3\cdot\bm\tau_1\times\bm\tau_2)
   [(\bm\sigma_3\cdot \bm\sigma_2\times\bm\sigma_1)y(r_{31})y(r_{23}) \nonumber \\
  &+ & (\bm\sigma_1\cdot \bm r_{31})
    (\bm\sigma_2\cdot \bm r_{23})(\bm\sigma_3\cdot\bm r_{31}\times \bm r_{23})
  t(r_{31})t(r_{23}) \nonumber \\
 &+ & (\bm\sigma_1\cdot \bm r_{31})(\bm\sigma_2\cdot \bm r_{31}\times\bm\sigma_3)
  t(r_{31})y(r_{23}) \nonumber \\
 &+ & (\bm\sigma_2\cdot \bm r_{32})(\bm\sigma_3\cdot \bm r_{32}\times
  \bm\sigma_1) y(r_{31})t(r_{23})] \nonumber \;\; ,
\end{eqnarray}
with $W_0$ an overall strength.
The $b$- and $d$-terms are present in the three models whereas the $a$-term
is present in the TM' and N2LOL and not in URIX. In the first two models,
the radial functions $y(r)$ and $t(r)$ are obtained from the following function
\begin{equation}
f_0(r)=\frac{12\pi}{m_\pi^3}\frac{1}{2\pi^2}
 \int_0^\infty dq q^2 \frac{j_0(qr)}{q^2+m_\pi^2}F_\Lambda(q)
\label{eq:f0r}
\end{equation}
where $m_\pi$ is the pion mass and
\begin{equation}
\left\{ \begin{array}{lll}
 y(r) & = & \frac{1}{r} f^\prime_0(r)   \\
 &\mbox{}& \\
 t(r) & = & \frac{1}{r} y^\prime(r)  \,\,\ .  
\end {array}
\right.
\label{eq:y0r}
\end{equation}
The cutoff function $F_\Lambda$
in the TM' or Brazil models is taken as
$[(\Lambda^2-m_\pi^2)/(\Lambda^2+q^2)]^2$. In the N2LOL model it is taken as
$\exp(-q^4/\Lambda^4)$. The momentum cutoff $\Lambda$ is a parameter of the model
fixing the scale of the problem in momentum space.
In the N2LOL, it has been taken $\Lambda=500$ MeV, whereas in the TM' model the quantity
$\Lambda/m_\pi$ has been varied to describe the triton or $^4$He binding energy
at fixed values of the constants $a$,$b$ and $d$. In the literature several cases
have been explored with typical
values around $\Lambda= 5 m_\pi$.

In the URIX model the radial dependence of the $b$- and $d$-terms is given in terms
of the functions
\begin{equation}
\left\{ \begin{array}{lll}
Y(r)& = &{\rm e}^{-x}/x\,\xi_Y   \\
 &\mbox{}& \\
T(r)& = &(1+3/x+3/x^2)Y(r)\,\xi_T
\end {array}
\right.
\label{eq:Y0r}
\end{equation}
with $x=m_\pi r$ and the cutoff functions are defined as
$\xi_Y=\xi_T=(1-{\rm e}^{-cr^2})$, with $c=2.1$ fm$^{-2}$.
This regularization has been used in the AV18 potential
as well. Since the parameters in the URIX model has been determined in
conjunction with the AV18 potential, the use of the same
regularization was a choice of consistency.
The relation between the functions $Y(r),T(r)$
and those of the previous models is
\begin{equation}
\left\{ \begin{array}{lll}
 Y(r) & =& y(r)+T(r)  \\
 &\mbox{}& \\
 T(r) & =& \frac{r^2}{3}t(r)\,\,\, .
\end {array}
\right.
\label{eq:T0r}
\end{equation}
With the definition given in Eq.(\ref{eq:f0r}), the asymptotic behaviour of
the functions $f_0(r)$, $y(r)$ and $t(r)$ is:
\begin{eqnarray}
&f_0(r\rightarrow\infty)&\rightarrow \frac{3}{m_\pi^2}\frac{{\rm e}^{-x}}{x} 
\nonumber \\
&y(r\rightarrow\infty)&\rightarrow -\frac{3{\rm e}^{-x}}{x^2}
       \left(1+\frac{1}{x}\right)  \\
&t(r\rightarrow\infty)&\rightarrow \frac{3}{r^2}\frac{{\rm e}^{-x}}{x}
       \left(1+\frac{3}{x}+\frac{3}{x^2}\right) \;\; .  \nonumber
\label{eq:f0rasymp}
\end{eqnarray}
In fact, with the normalization chosen for $f_0$, the functions $Y$ and
$T$ defined from $y$ and $t$ in Eq.~(\ref{eq:T0r}) and those ones defined 
in the URIX model in Eq.~(\ref{eq:Y0r})
coincide at large separation distances. Conversely, they have a
different short range behavior.

The last two terms in Eq.~(\ref{eq:w123}) correspond to a 2N contact term
with a pion emitted or absorbed ($D$-term) and to a 3N contact interaction
($E$-term). Their local form, in configuration space,
derived from Ref.~\cite{N2LO}, are
\begin{eqnarray}
W_D(1,2,3) & = & W_0^D (\bm\tau_1\cdot\bm\tau_2) \times  \nonumber \\
 & \{ & \!\! (\bm\sigma_1\cdot\bm\sigma_2)
  [y(r_{31})Z_0(r_{23})+Z_0(r_{31})y(r_{23})] \nonumber  \\
 & + & (\bm\sigma_1\cdot \bm r_{31})(\bm\sigma_2\cdot \bm r_{31})
  t(r_{31})Z_0(r_{23}) \nonumber \\
 & + &(\bm\sigma_1\cdot \bm r_{32})(\bm\sigma_2\cdot \bm r_{32})
  Z_0(r_{31})t(r_{23})\} \\
W_E(1,2,3) & = & W_0^E(\bm\tau_1\cdot\bm\tau_2) Z_0(r_{31})Z_0(r_{23})  \,\, .
\nonumber 
\end{eqnarray}
The constant $W_0^D,W_0^E$ fix the strength of these terms.
In the case of the URIX model the $E$-term is present without the isospin operator
structure and it has been included as purely phenomenological, without
justifying its form from a particular exchange mechanism. Its radial dependence
has been taken as $Z_0(r)=T^2(r)$.
In the N2LOL model, the function $Z_0(r)$ is defined as
\begin{equation}
Z_0(r)=\frac{12\pi}{m_\pi^3}\frac{1}{2\pi^2}
 \int_0^\infty dq q^2 j_0(qr) F_\Lambda(q)
\label{eq:z0r}
\end{equation}
with the same cutoff function used in the definition of $f_0$ in Eq.(~\ref{eq:f0r}), 
$F_\Lambda(q)=\exp(-q^4/\Lambda^4)$.
In the TM' model the $D$- and $E$-terms are absent.

Each model is now identified from the values assigned to the different
constants $a,b,d,c_D,c_E$.
Following Refs.~\cite{tmp,nogga02}, in the case of the TM' model,
the values of the constants have been chosen as $a=-0.87\; m^{-1}_\pi$,
$b=-2.58\; m^{-3}_\pi$, and $d=-0.753\; m^{-3}_\pi$; the strength
$W_0=(gm_\pi/8\pi m_N)^2\;m_\pi^4$ and
the cutoff has been fixed to $\Lambda=4.756\;m_\pi$ in order to describe
correctly $B$($^4$He).
In Table~\ref{tb:table1} the calculations have been
done using these values with $g^2=197.7$, $m_\pi=139.6$ MeV,
$m_N/m_\pi=6.726$ ($m_N$ is the nucleon mass)
as given in the original derivation of the TM potential.
As mentioned before, this model does not include the $D$- and $E$-terms.

In the URIX model the $b$- and $d$-terms are present, however with
a fix relative value. The strength of these terms is:
$bW_0=4\;A^{PW}_{2\pi}$ and $d=b/4$, with $A^{PW}_{2\pi}=-0.0293$ MeV.
The model includes a purely central repulsive term introduced to
compensate the attraction of the previous term, which
by itself would produce a large overbinding in infinite nuclear matter.
It is defined as
\begin{equation}
  W_E^{URIX}(1,2,3)=A_R T^2(r_{31})T^2(r_{23})
\end{equation}
with $A_R=0.0048$ MeV. 

In the N2LOL potential the constants of the
$a$-, $b$-, $d$-, $D$- and $E$-terms are defined in the following way:
\begin{eqnarray}
&  W_0=\frac{1}{12\pi^2}\left(\frac{m_\pi}{F_\pi}\right)^4g^2_A m_\pi^2 \nonumber \\
&  W_D^0=\frac{1}{12\pi^2}\left(\frac{m_\pi}{F_\pi}\right)^4
        \left(\frac{m_\pi}{\Lambda_x}\right) \frac{g_A m_\pi}{8}  \\
&  W_E^0=\frac{1}{12\pi^2}\left(\frac{m_\pi}{F_\pi}\right)^4
        \left(\frac{m_\pi}{\Lambda_x}\right) m_\pi \nonumber
\end{eqnarray}
with $a= c_1 m^2_\pi$, $b= c_3/2$, $d= c_4/4$,  and
$c_1=-0.00081$ MeV$^{-1}$, $c_3=-0.0032$ MeV$^{-1}$, $c_4=-0.0054$ MeV$^{-1}$
taken from Ref.~\cite{entem}. The other two constants, $c_D=1.0$ and $c_E=-0.029$,
have been determined in Ref.~\cite{N2LO} from a fit to $B$($^3$H)
and $B$($^4$He) using the N3LO-Idaho+N2LOL potential model.
The numerical values of the constant entering in
$W_0$, $W^0_D$ and $W^0_E$ are taken as $m_\pi=138$ MeV, $F_\pi=92.4$ MeV,
$g_A=1.29$, and the chiral symmetry breaking scale $\Lambda_x=700$ MeV.

\begin{figure*}[!hbt]
\begin{center}
%\vspace{1cm}
\includegraphics[scale=0.6,angle=0]{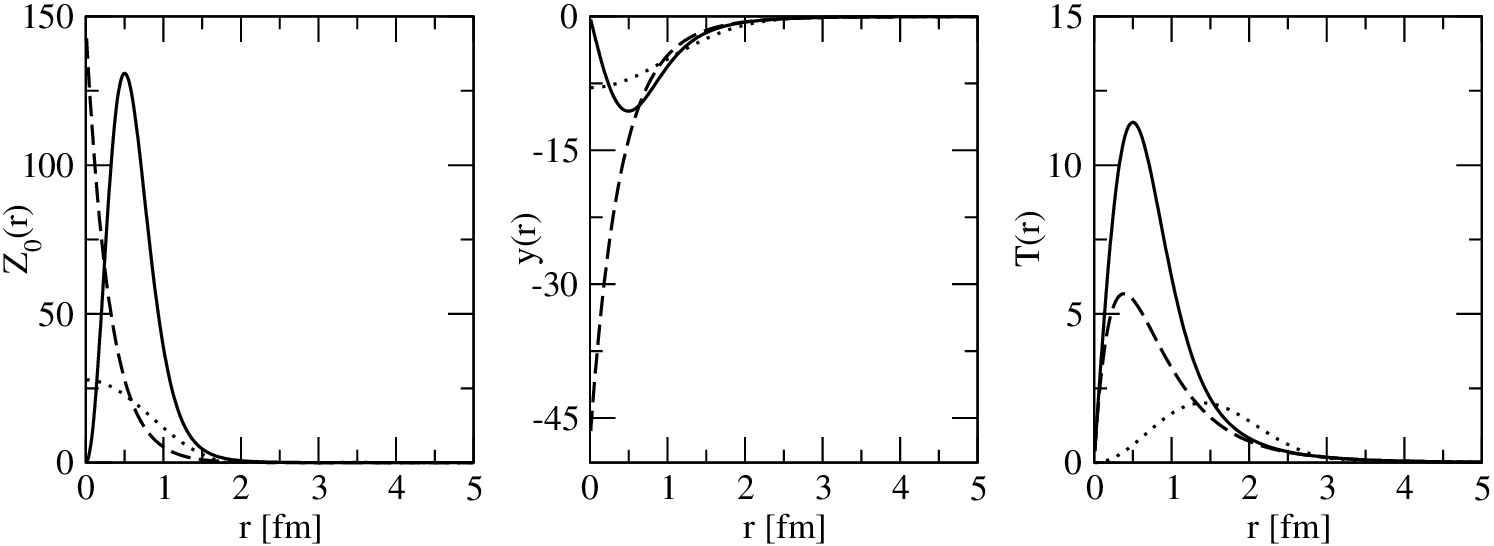}
\caption{ The $Z_0(r)$, $y(r)$ and $T(r)$ functions as functions of the
interparticle distance $r$ for the URIX (solid line), TM' (dashed line) and
N2LOL (dotted line) models.}
\label{fig:functions}
\end{center}
\end{figure*}

In order to analyze the different short range structure of the TNF models,
in Fig.~\ref{fig:functions} we compare the non-dimensional functions
$Z_0(r)$, $y(r)$ and $T(r)$ for the three models under consideration.
In the TM' model using the definition
of Eq.(\ref{eq:z0r}) and using the corresponding cutoff function we can define:

\begin{eqnarray}
 Z^{TM}_0(r) & = & \frac{12\pi}{m_\pi^3}\frac{1}{2\pi^2}
 \int_0^\infty dq q^2 j_0(qr) \left(\frac{\Lambda^2-m_\pi^2}{\Lambda^2+q^2}
\right)^2  \nonumber \\
 & = & \frac{3}{2}\left(\frac{m_\pi}{\Lambda}\right)
   \left(\frac{\Lambda^2}{m_\pi^2}-1\right)^2 {\rm e}^{-\Lambda r}  \;\; .
\label{eq:z0rtm}
\end{eqnarray}
This function is showed in the first panel of Fig.~\ref{fig:functions}
as a dashed line.
From the figure we can see that, in the case of the URIX model, the functions
$Z_0(r)$ and $y(r)$ go to zero as $r\rightarrow 0$. This is not the case for the
other two models and is a consequence of the regularization choice of the $Y$
and $T$ functions adopted in the URIX.

\section{Parametrization Study of the Three Nucleon Forces}
\label{kievsky_sec:2}

In this section we study possible variations to the parametrization of
the TNF models in order to describe the $A=3,4$ binding energies and
$^2a_{nd}$.

\subsection{Tucson-Melbourne Force}
\label{kievsky_subsec2:1}

\begin{figure*}[!htb]
\begin{center}
\vspace{1.5cm}
\includegraphics[scale=0.6,angle=0]{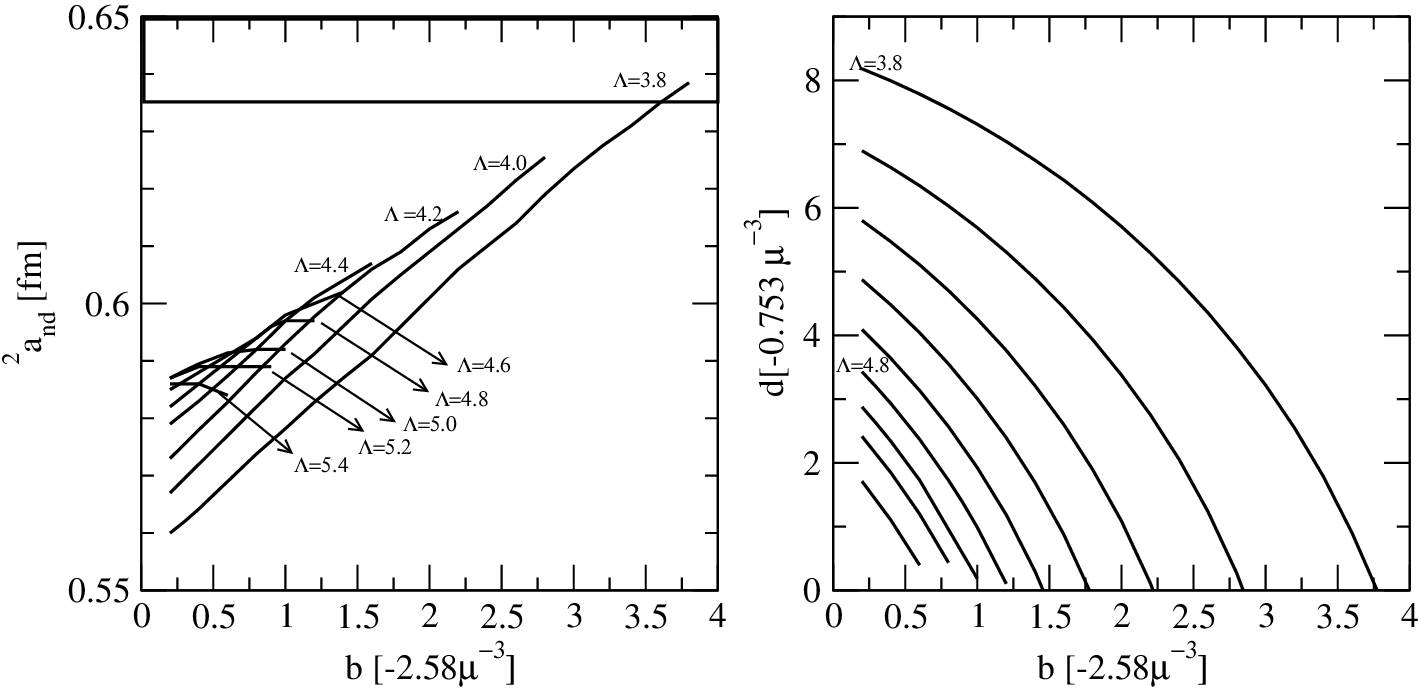}
\caption{ The doublet scattering length $a_{n-d}$ as a function of
the parameter $b$ of the TM' potential (right panel) for different values
of the cutoff. The corresponding values of the parameter $d$ used to
reproduce the triton binding energy (left panel).}
\label{fig:tucson2}
\end{center}
\end{figure*}

We first study the TM' potential and
we would like to see if, using the AV18+TM' interaction, it is possible
to reproduce simultaneously the triton binding energy and the
doublet $n-d$ scattering length for some values of the parameters.
The $a$-term gives a very small contribution to these quantities,
therefore, in the following analysis we maintain it fixed at the value
$a=-0.87\; m^{-1}_\pi$.
In Fig.~\ref{fig:tucson2}, left panel, the doublet $n-d$ scattering length
is given as a function of the parameter $b$ (in units of
its original value $b=-2.58\; m^{-3}_\pi$) for
different values of the cutoff $\Lambda$ (in units of $m_\pi$). The box
in the figure includes values compatible with the experimental results.
The value of the constant $d$ has been fixed to reproduce the triton
binding energy. The corresponding values of
the parameter $d$ (in units of its original value $d=-0.753\; m^{-3}_\pi$)
are given in the right panel as a function of $b$.
Each point of the curves in both panels
corresponds to a set of parameters that, in connection with the AV18 potential,
reproduces the triton binding energy.
The variations of the parameters given in Fig.~\ref{fig:tucson2} do not
exhaust all the possibilities. However we can observe that,
with the AV18+TM' potential, there is a very small region in the parameter's
phase space available for a simultaneous description of
the triton binding energy and the doublet scattering length. 
This small region corresponds to a big value of $b$ and $d$ results to be
almost zero. Moreover, the value of the cutoff $\Lambda$ around $3.8m_\pi$ is smaller
than the values usually used with the TM' potential ($\Lambda\approx5m_\pi$).

To be noticed that, for negative values of
the parameters $a$, $b$ and $d$, the TM' potential is attractive. It
does not include explicitly a repulsive term. Added to a specific NN
potential that underpredicts the three-nucleon binding energy,
it supplies the extra binding by fixing appropriately its strength.
However, as mentioned in the Introduction, the scattering length is sensitive to
the balance between the attractive part and the repulsive part of the
complete interaction. Therefore, it seems that supplying only an attraction,
fixed to reproduce the triton binding energy, in the case of the TM'
interaction it is difficult to reproduce correctly this balance.

\begin{figure*}[!htb]
\begin{center}
\vspace{1cm}
\includegraphics[scale=0.6,angle=0]{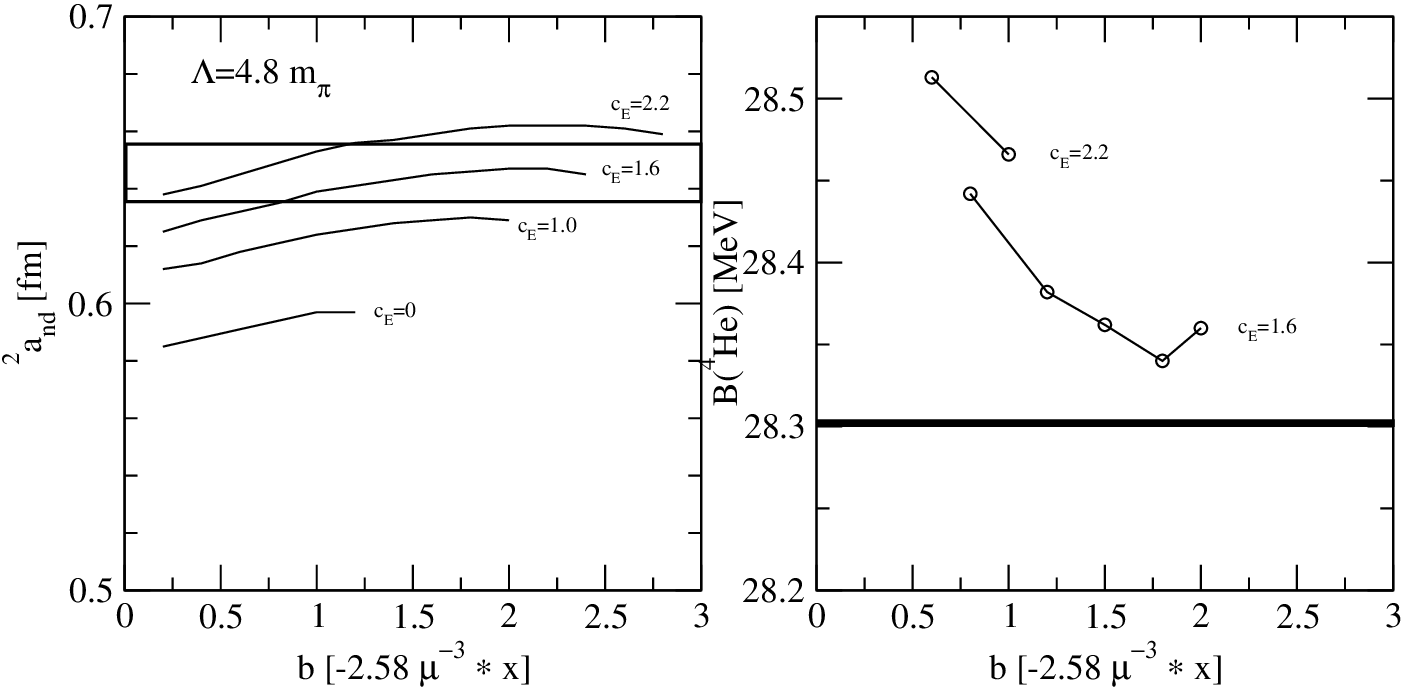}
\caption{ The doublet scattering length $a_{n-d}$ as a function of
the parameter $b$ of the TM' potential including the $W_E^{TM}$-term,
for different values
of the strength $c_E$ (right panel). The corresponding values of $B(^4{\rm He})$
(left panel).}
\label{fig:tucson3}
\end{center}
\end{figure*}

As discussed before, the TM' potential is a modification of the original
TM potential compatible with chiral symmetry. At the same order
(next-to-next-to-leading order) in the chiral effective field theory
the $D$- and $E$-terms appear (see Ref.~\cite{epelbaum02} and references therein)
as given in Eq.(\ref{eq:w123}).
Here we introduce the following additional term to the TM' potential
based on a contact term of three nucleons
\begin{equation}
W^{TM}_E(1,2,3)=W^0_E\sum_{cyc}Z^{TM}_0(r_{31})Z^{TM}_0(r_{23})  \,\, .
\end{equation}
This term is similar to the repulsive term of the URIX model and, 
for the sake of simplicity,
we do not include the $({\bm \tau}_1\cdot{\bm \tau}_2)$ operator.
The function $Z_0^{TM}$ is a positive function, therefore, for positive values of
$c_E$, the new term is repulsive. We include it in the following
analysis of the TM' potential.
The analysis of the new term is given in Fig.~\ref{fig:tucson3}. In the left panel
the doublet $n-d$ scattering length
is given as a function of the parameter $b$ (in units of
its original value $b=-2.58\; m^{-3}_\pi$) for
different values of the strength of the $W_E^{TM}$-term. The value of the
cutoff $\Lambda$ has been fixed to $4.8\;m_\pi$. The box
in the figure includes values compatible with the experimental results.
Moreover, the value of the constant $d$ has been fixed to reproduce the triton
binding energy. The corresponding values of the $^4$He binding energy,
$B(^4{\rm He})$, is given in the right panel.

Comparing the left panels in Figs.~\ref{fig:tucson2} and~\ref{fig:tucson3}, the
effect of the new term is clear. In Fig.~\ref{fig:tucson2} we see that using
$\Lambda=4.8\;m_\pi$, $^2a_{nd}$ is not well reproduced. Conversely,
in Fig.~\ref{fig:tucson3}, the inclusion of the new term moves this curve in
the correct direction and with values of its strength around $c_E=1.6$ it is
possible to reproduce the experimental value of $^2a_{nd}$. There is also an
improvement in the description of $B(^4{\rm He})$. In fact, the AV18+TM'
model with $\Lambda=4.8\;m_\pi$ reproduces the triton binding energy
as can be seen from Fig.~\ref{fig:tucson2}. However it predicts
$B(^4{\rm He})=28.55$ MeV, which is slightly too high. With the $W_E^{TM}$-term,
at $c_E=1.6$, the description of $B(^4{\rm He})$ improves. For example with
$b=-3.87\; m^{-3}_\pi$, $d=-3.375\; m^{-3}_\pi$ and $\Lambda=4.8\;m_\pi$,
we obtain $B(^4{\rm He})=28.36$ MeV, very close to the experimental value. 

\subsection{UrbanaIX Force}
\label{kievsky_subsec2:2}

\begin{figure*}[!htb]
\begin{center}
\vspace{1.5cm}
\includegraphics[scale=0.6,angle=0]{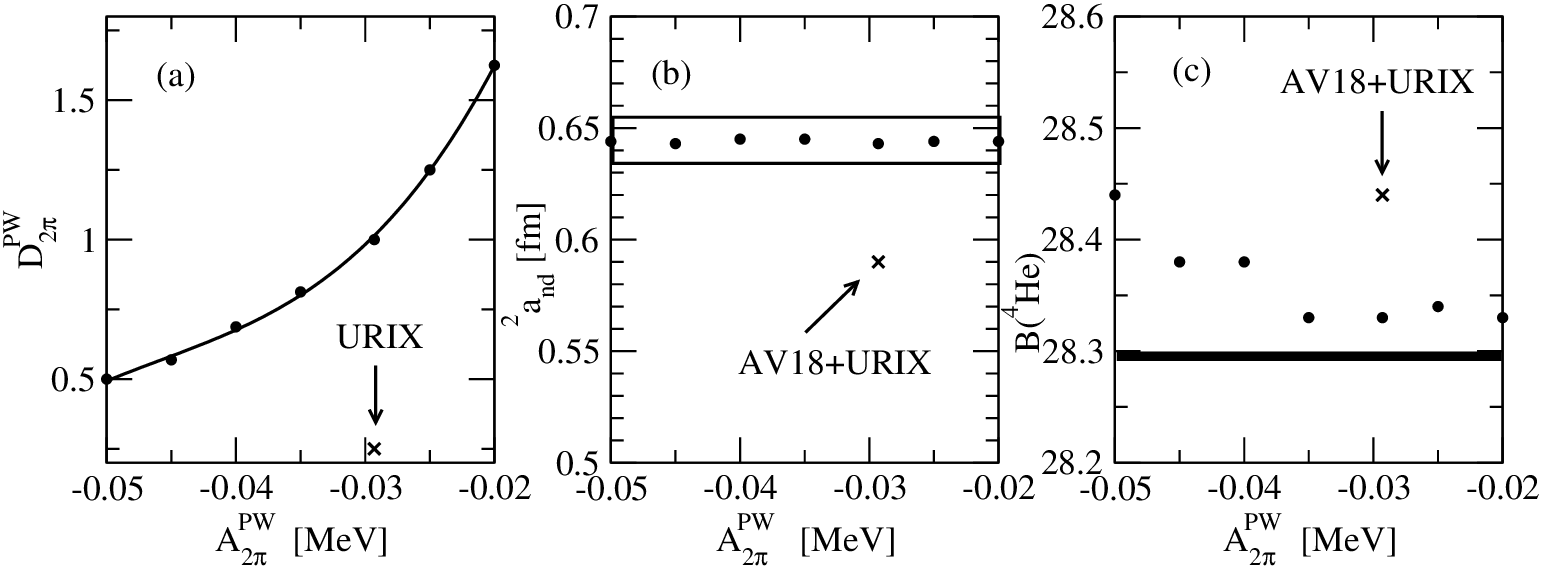}
\caption{ (a) The relative strength $D^{PW}_{2\pi}$ as a function of $A^{PW}_{2\pi}$.
In each point of the curve the triton binding energy and $^2a_{nd}$ are
well described. (b) Values of $^2a_{nd}$ for the seven combinations of the parameters
indicated as solid points in panel (a). (c) The corresponding predictions for 
$B$($^4$He). The crosses indicate the results using the parameters defined in the URIX
model}

\label{fig:urbana1}
\end{center}
\end{figure*}

In the following we analyze the URIX potential which has two parameters,
$A^{PW}_{2\pi}$ and $A_R$.
In this model the strength of the $d$-term was related to the strength of
the $b$-term as $b=4d$. The original values of the parameters were fixed in
Ref.~\cite{urbana} in conjunction with the AV18 NN potential and,
from Table~\ref{tb:table1}, we observe that the model correctly describes
the triton binding energy. However, it overestimates $B$($^4$He) and underestimates
$^2a_{nd}$. In order to improve the description of these quantities,
we have varied the constants $A^{PW}_{2\pi}$, $A_R$ and
the relative strength $D^{PW}_{2\pi}=d/b$ between the $b$- and $d$-terms.
For a given value of $A^{PW}_{2\pi}$, the values of
$A_R$ and $D^{PW}_{2\pi}$ has been chosen to reproduce $B(^3{\rm H})$ and
$^2a_{nd}$. The results are given in Fig.~\ref{fig:urbana1}. In panel (a),
$A^{PW}_{2\pi}$ is given as a function of $D^{PW}_{2\pi}$ with $A_R$
varying from $0.0176$ MeV at $A^{PW}_{2\pi}=-0.02$ to
$0.0210$ MeV at $A^{PW}_{2\pi}=-0.050$ MeV. These values of $A_R$ are
more than three times greater than the original value. In panel
(b) and (c) the results for $^2a_{nd}$ and $B(^4{\rm He})$ are given
respectively. The latter has not been included in the determination
of the parameters, however we observe a rather good description in particular 
for values of $D^{PW}_{2\pi}>0.7$.

With a modification of the parameters in the URIX force, we were able to describe
$B$($^3$H), $^2a_{nd}$ and $B$($^4$He). This has been achieved with a
substantial increase of the repulsive term. Also $D^{PW}_{2\pi}$ is
quite far from its original value. For example,
at the original value of $A^{PW}_{2\pi}=-0.0293$ MeV,
the relative strength is $D^{PW}_{2\pi}=1$ and  $A_R=0.0181$ MeV. This is
four times and more than three of the original values, respectively. As
$D^{PW}_{2\pi}$ diminishes, $A_R$ tends to increase further with the consequence
that the mean value of the repulsive part of $W$ results to be more than three times
the original AV18+URIX value. This is compensated by a lower mean value of the kinetic 
energy.
A further analysis of the effects of the new parametrizations 
is done in the next section studying selected $p-d$ polarization
observables.

\subsection{N2LOL Force}

\begin{figure}[htb]
\begin{center}
\vspace{2cm}
\includegraphics[scale=0.6,angle=0]{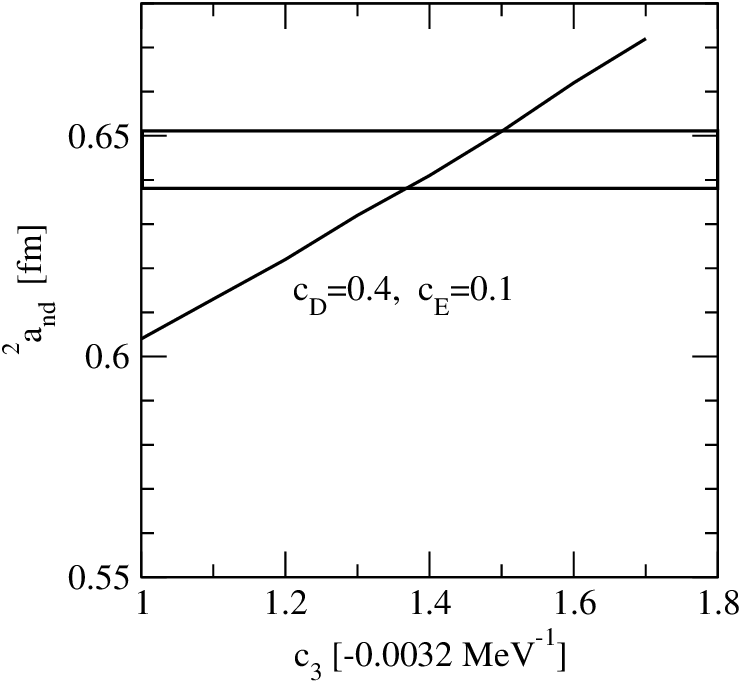}
\caption{ $^2a_{nd}$ as a function of the $c_3$ parameter in the N2LOL model.}
\end{center}
\label{fig:n2lo1}
\end{figure}

The parameters $c_1$, $c_3$ and $c_4$ of the N2LOL have been taken from
the chiral N3LO NN force of Ref.~\cite{entem}, whereas the $c_D$ and $c_E$
parameters have been determined in Ref.~\cite{N2LO}, in conjunction with
that NN force, by fitting $B$($^3$H) and
$B$($^4$He). Here we are going to use the N2LOL force in conjunction with
the AV18 NN interaction, so we have to modify its parametrization since
the amount of attraction to be gained is now different
(see Table~\ref{tb:table1}). Moreover, the modification has to be done in such
a way that $B$($^3$H) and $^2a_{nd}$ are well reproduced. As an example,
in Fig.~\ref{fig:n2lo1}, $^2a_{nd}$ is shown as a function of the parameter
$c_3$ (in units of its original value $c_3=0.0032$ MeV$^{-1}$) fixing
$c_D=0.4,c_E=0.1$ and varying $c_4$ in order to reproduce $B$($^3$H).
With the values $c_3=-0.0048$ MeV$^{-1}$, $c_4=0.0043$ MeV$^{-1}$,
$^2a_{nd}$ fall inside the box and matches the experimental value.
In this case, the$^4$H binding energy results $B(^4{\rm H})=28.36$ MeV.

\section{Polarization observables with the new parametrizations}

\begin{figure*}[htb]
\begin{center}
\vspace{2cm}
\includegraphics[scale=0.7,angle=0]{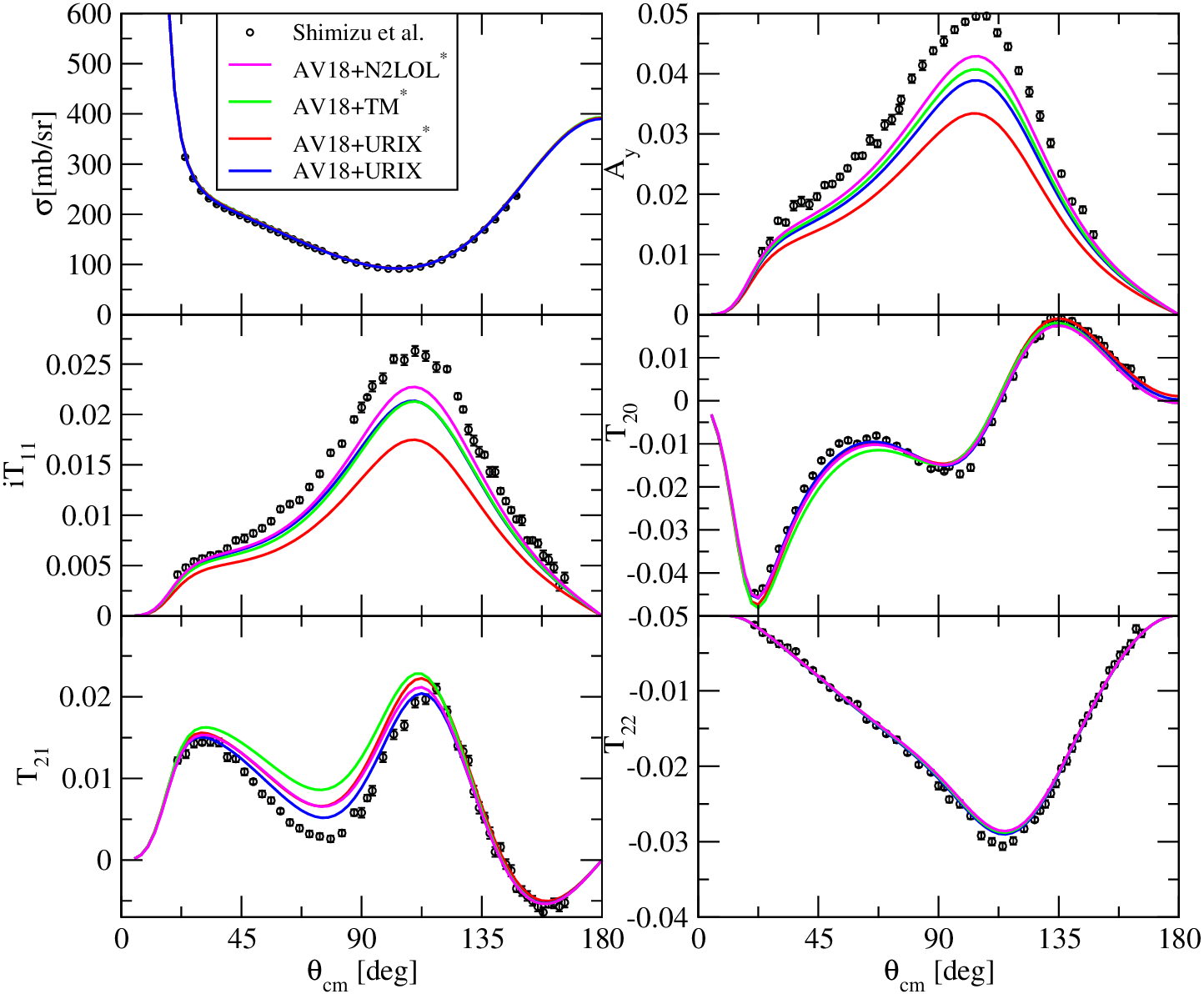}
\caption{ Cross section, vector and tensor analyzing powers for $p-d$
scattering at $E_{lab}=3$ MeV. Experimental points are for Ref.~\protect\cite{shimizu}}
\end{center}
\label{fig:all1}
\end{figure*}

In the previous section we have analyzed different parametrizations of the
TM', URIX and N2LOL TNFs determined in conjunction with the AV18 NN potential.
With the new parametrizations the three quantities under observation,
$B$($^3$H), $^2a_{nd}$ and $B(^4{\rm He})$, are well reproduced. However,
some substantial modifications to the first two models were necessary.
In the case of the TM' interaction, we found necessary to include a repulsive term.
In the analysis of the URIX interaction, the strength of the repulsive
term resulted to be more than three times larger. In the case of the
N2LOL interaction, a minor adjustment of the parameters was necessary. Now we would
like to analyze the effects of the new parametrizations in observables that
are not correlated to the binding energies or to $^2a_{nd}$. Some polarization
observables in $p-d$ scattering have this characteristic, in particular the
vector and tensor analyzing powers.
In Fig.~\ref{fig:all1}, the differential cross section $d\sigma/d\Omega$, 
the vector polarization observables $A_y$ and $iT_{11}$ and the tensor polarization 
observables $T_{20}$, $T_{21}$ and $T_{22}$ are shown at the laboratory energy
$E_{lab}=3$ MeV, for the different potential
models. As a reference we use the AV18+URIX interaction given in the figure 
as a blue line. 
In the figure, the other three curves corresponds to particular parametrizations
of the models that reproduce $^2a_{nd}$ and $B$($^3$H) and approximate, as much
as possible, $B(^4{\rm He})$. The parametrizations of the models selected for the figure
are the following: the AV18+URIX$^*$ model is defined with 
$A_{2\pi}^{PW}=-0.0293$ MeV, $D_{2\pi}^{PW}=1$ and $A_R=0.018$ MeV.
In the AV18+TM$^*$ model we have used
$a=-0.87\; m^{-1}_\pi$, $b=-9.804\; m^{-3}_\pi$, $d=-3.1657\; m^{-3}_\pi$,
$c_E=1$, and $\Lambda=4 m_\pi$.
In the AV18+N2LO$^*$ model the parametrization corresponds to
$c_1=-0.00081$ MeV$^{-1}$ (its original value),
$c_3=-0.0048$ MeV$^{-1}$, $c_4=-0.0043$ MeV$^{-1}$, $c_D=0.4$ and $c_E=0.1$.
From the figure we can observe that the models
describe equally well the differential cross section and the tensor analyzing powers
$T_{20},T_{22}$. Differences are observed in the vector analyzing powers $A_y$ and
$iT_{11}$. Taking as a reference the results of the AV18+URIX model,
in both cases the AV18+URIX$^*$ model produces a noticeable worse description
whereas the AV18+N2LOL$^*$ slightly improves the description. The new
parametrizations of the TNF models overpredict $T_{21}$ in all cases, 
in particular the AV18+TM$^*$ model.

\section{The Kohn Variational Principle in terms of Integral Relations}
Recently two integral relations have been derived from the KVP~\cite{intrel}.
It has been shown that starting from the KVP, the tangent of the phase-shift
can be expressed in a form of a quotient where both, the numerator and the
denominator, are given as two integral relations. 
Let us first consider a two-body system interacting
through a short-range potential $V(r)$ at the center of mass energy $E$
in a relative angular momentum state $l=0$.
The solution of the Schr\"odinger equation in configuration space
($m$ is twice the reduced mass),
\begin{equation}
(-\frac{\hbar^2}{m}\nabla^2+V-E)\Psi(\bm r)=0 \;\; ,
\end{equation}
can be obtained after specifying the corresponding boundary conditions.
For $E>0$, with $k^2=E/(\hbar^2/m)$ and assuming a short-range potential $V$,
$\Psi(\bm r)=\phi(r)/\sqrt{4\pi}$ and
\begin{equation}
\phi(r\rightarrow\infty)\longrightarrow
\sqrt{k} \left[A\frac{\sin(kr)}{kr}+B\frac{\cos(kr)}{kr}\right ] \;\; .
\end{equation}
With the above normalization, the solution $\Psi$ verifies the following
integral relations:
\begin{eqnarray}
-&\frac{m}{\hbar^2} <\Psi|H-E|F>=B
 &{\rm with} \hspace{0.5cm} F=\sqrt{\frac{k}{4\pi}} \frac{\sin(kr)}{kr} \cr
&\frac{m}{\hbar^2} <\Psi|H-E|G>=A
 &{\rm with} \hspace{0.5cm} G=\sqrt{\frac{k}{4\pi}} \frac{\cos(kr)}{kr}  \cr
& \tan\delta = \frac{B}{A} \;\; .
\label{rel1}
\end{eqnarray}
Explicitly they are
\begin{eqnarray}
&-& \frac{m}{\hbar^2\sqrt{k}}\int_0^\infty dr
\sin(kr)V(r)[r\phi(r)]=B  \cr
& & \cr
& & \frac{m}{\hbar^2\sqrt{k}}\int_0^\infty dr
\cos(kr)V(r)[r\phi(r)]+\frac{\phi(0)}{\sqrt{k}} =A,
\label{origin}
\end{eqnarray}
where in the last integral we have used the property
$\nabla^2(1/r)=-4\pi\delta({\bm r})$.

In practical cases the solution of the Schr\"odinger equation is obtained
numerically. Then, $\tan\delta$ is extracted from $\phi(r)$ analyzing its
behavior
outside the range of the potential. The equivalence between the extracted
value and that one obtained from the integral relations defines the accuracy
of the numerical computation. A relative difference of the order of
$10^{-7}$ of the two values is usually achieved using standard numerical
techniques to solve the differential equation and to compute the two
one-dimensional integrals. To be noticed the short range character of the
integral relations. This means that the phase-shift is determined by
the internal structure of the wave function.

The last relation in Eq.~\refeq{origin} shows a dependence on the value of the
wave function at the origin. It could be convenient to eliminate this explicit
dependence since the numerical determination of $\phi(0)$ might be problematic,
as we will show. To this end we introduce a regularized function
$\tilde G=f_{reg}G$ with the property $|\tilde G(r=0)|<\infty$ and
$\tilde G=G$ outside the interaction region. A possible choice is
\begin{equation}
 \tilde G=\sqrt{\frac{k}{4\pi}}\frac{\cos(kr)}{kr}(1-{\rm e}^{-\gamma r})\;\; ,
\end{equation}
where the regularization function $f_{reg}=(1-{\rm e}^{-\gamma r})$ has been
introduced
with $\gamma$ being a non linear parameter which will be discussed below.
Values verifying $\gamma>1/r_0$,
with $r_0$ the range of the potential, could be appropriate.
The regularized function $\tilde G$ (as well as the irregular function $G$),
verifies the normalization condition
\begin{equation}
\frac{m}{\hbar^2}\left[<F|H-E|\tilde G>-<\tilde G|H-E|F>\right] =1 \;\; .
\label{norm}
\end{equation}
Therefore the second integral relation in Eq.~\refeq{rel1} remains valid using
$\tilde G$ in place of $G$,
\begin{equation}
 \frac{m}{\hbar^2} <\Psi|H-E|\tilde G>=A \;\; ,
\label{reg1}
\end{equation}
with the explicit form:
\begin{equation}
  \frac{m}{\hbar^2\sqrt{k}}\int_0^\infty d{r}
 \cos(kr)V(r)[r\phi(r)]+ I_\gamma =A
\label{reg2}
\end{equation}
where in $I_\gamma$ all terms depending on $\gamma$, introduced by
$f_{reg}$, are included. Comparing Eq.~\refeq{reg2} to Eq.~\refeq{origin} 
we identify $I_\gamma=\phi(0)/\sqrt{k}$.

In the following we demonstrate that the relation $\tan\delta=B/A$, which is
an exact relation when the exact wave function $\Psi$ is used in Eq.~\refeq{rel1},
can be considered accurate up to second order when a trial wave function is
used, as it has a strict connection with the Kohn variational principle.

The connection of the integral relations with the KVP is straightforward.
Defining a trial wave function $\Psi_t$ as
\begin{equation}
 \Psi_t=\Psi_c +AF+B\;\tilde G  \;\; ,
\label{psic}
\end{equation}
with $\Psi_c\rightarrow 0$ as
$r\rightarrow\infty$, the condition
$\Psi_t\rightarrow A F+B \; G$ as $r\rightarrow\infty$ is fulfilled. The KVP
states that the second order estimate for $\tan\delta$ is
\begin{equation}
[\tan\delta]^{2^{nd}}=\tan\delta - \frac{m}{\hbar^2}<(1/A)\Psi_t|H-E|(1/A)\Psi_t> \,\ .
\label{kohnn}
\end{equation}
The above functional is stationary with respect to variations
on $\Psi_c$ and $\tan\delta$. Without loosing generality $\Psi_c$ can be
expanded in a (square integrable) complete basis
\begin{equation}
 \Psi_c=\sum_n a_n \phi_n(r) \;\; .
\end{equation}
The variation of the functional with respect to the linear
parameters $a_n$
and $\tan\delta$ leads to the following equations
\begin{eqnarray}
 &<\phi_n|H-E|\Psi_t>=0  \cr
 &  \cr
 &<\tilde G|H-E|\Psi_t>=0  \;\;\; .
\label{first}
\end{eqnarray}
To obtain the last equation, the normalization relation of Eq.~\refeq{norm}
has been used.
From these two equations, $\Psi_c$ and the first order
estimate of the phase shift $(\tan\delta)^{1^{st}}$ can be determined.
To be noticed that the first equation implies
$<\Psi_c|H-E|\Psi_t>=0$. Furthermore, from the general relation
$(m/{\hbar^2})\left[<\Psi_t|H-E|\tilde G>-
<\tilde G|H-E|\Psi_t>\right]=A$, and using
the second equation in Eq.~\refeq{first}, the following integral relation results

\begin{equation}
\frac{m}{\hbar^2}<\Psi_t|H-E|\tilde G>=A  \;\; .
\end{equation}

Replacing the two relations of Eq.\refeq{first} into the functional of
Eq.\refeq{kohnn}, a second order estimate of the phase shift is obtained
\begin{equation}
[\tan\delta]^{2^{nd}}=(\tan\delta)^{1^{st}}
- \frac{m}{\hbar^2}<F|H-E|(1/A)\Psi_t> \,\ .
\label{second1}
\end{equation}
Multiplying Eq.~\refeq{second1} by $A$ one gets
\begin{equation}
B^{2^{nd}}=B^{1^{st}}
- \frac{m}{\hbar^2}<F|H-E|\Psi_t> \,\ .
\label{second2}
\end{equation}
On the other hand, a first order estimate for the coefficient $B$ can be obtained
from the general relation
\begin{equation}
\frac{m}{\hbar^2}\left[<F|H-E|\Psi_t>-
           <\Psi_t|H-E|F>\right]=B^{1^{st}} \,\,\ .
\label{firstb}
\end{equation}
Therefore,
replacing Eq.\refeq{firstb} in Eq.\refeq{second2}, a second order
integral relation for $B$ is obtained. The above results can be
summarized as follow
\begin{eqnarray}
 B^{2^{nd}}& = & -\frac{m}{\hbar^2}<\Psi_t|H-E|F>   \cr
 && \cr
 A & = & \frac{m}{\hbar^2}<\Psi_t|H-E|\tilde G>  \cr
 && \cr
 [\tan\delta]^{2^{nd}} & = & B^{2^{nd}}/A  \,\, .
\label{relint}
\end{eqnarray}

These equations extend the validity of the integral relations,
given in Eq.\refeq{rel1} for the exact wave functions, to trial
wave functions. To be noticed that $F,\tilde G$ are solutions
of the Schr\"odinger equation in the asymptotic region, therefore
$(H-E)F\rightarrow 0$ and $(H-E)\tilde G\rightarrow 0$
as the distance between the particles increases.
As a consequence the decomposition of $\Psi_t$ in
the three terms of Eq.~\refeq{psic} can be considered formal since,
due to the short-range character
of the relation integrals, it is sufficient that the trial wave function
be a solution of $(H-E)\Psi_t=0$ in the interaction region, without
an explicit indication of its asymptotic behavior. This fact,
together with the variational character
of the relations allows for a number of applications to be discussed
in the next sections.

\section{Integral Relations for $A=2,3$ systems}

Applications of the integral relations to systems with $A=2,3$
are given. We first consider the following central, 
$s$-wave gaussian potential
\begin{equation}
 V(r)=-V_0\exp{(-r^2/r_0^2)}  \;\; ,
\end{equation}
with $V_0=-51.5$ MeV, $r_0=1.6$ fm and $\hbar^2/m=41.4696$ MeV fm$^2$.
This potential has a shallow $L=0$ bound state with energy
$E_{2B}=-0.397743$ MeV.

In the $A=2$ system, the orthogonal basis
\begin{equation}
 \phi_m={\cal L}_m^{(2)}(z)\exp{-(z/2)}  \;\; ,
\end{equation}
with ${\cal L}_m$ a (normalized)
Laguerre polynomial and $z=\beta r$, being $\beta$
a nonlinear parameter, is used to expand the wave function of the system
\begin{equation}
 \Psi_0=\sum_{m=0}^{M-1} a^0_m \phi_m  \,\, .
\end{equation}
The Schr\"odinger equation is transformed to an eigenvalue problem that can
be solved for different values of the dimension $M$ of the
basis. The variational principle states that
\begin{equation}
E_0=\bra \Psi_0|H|\Psi_0 \ket \ge E_{2B}  \;\; ,
\end{equation}
with the equality obtained for $M\rightarrow\infty$.
The nonlinear parameter $\beta$ can be fixed to make improve the
convergence properties of the basis. In fact,
for each value of $M$ there is a value of $\beta$ that minimizes the energy.
Increasing $M$, the minimum of the energy becomes less dependent
on $\beta$ resulting in a plateau.
Increasing further the dimension of the basis, the extension
of the plateau increases as well, without any appreciable improvement in
the eigenvalue, indicating that the convergence has been reached up to
certain accuracy. At each step $\Psi_0$
represents a first order estimate of the bound state exact wave function.

\begin{figure}
\begin{center}
\vspace{1.0cm}
\includegraphics[scale=0.35,angle=0]{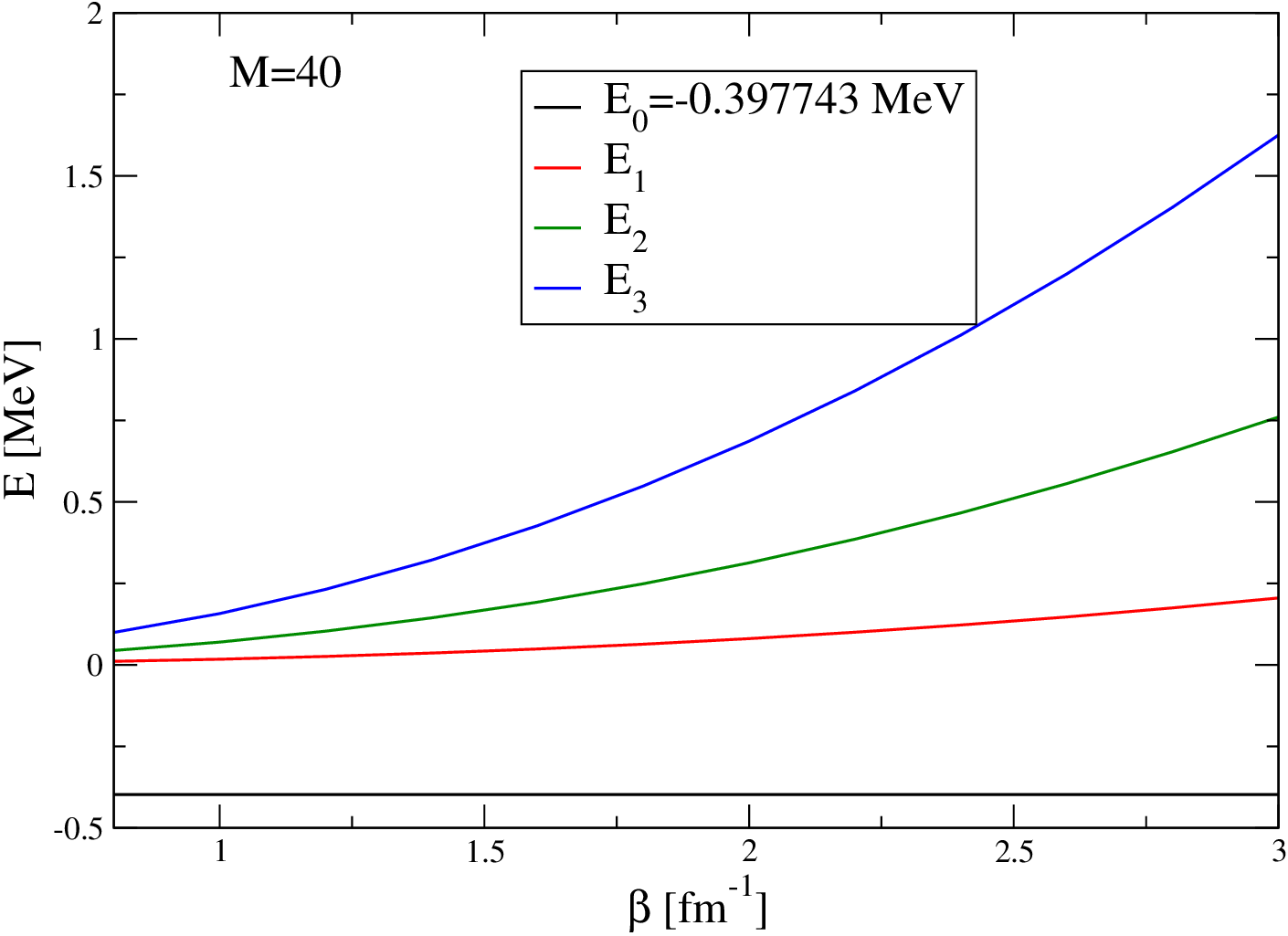}
\caption{The two-nucleon bound state energy $E_0$ and the first three positive
 eigenvalues $E_j$ as a function of $\beta$ in the case of $M=40$}
\label{fig:itfig1}
\end{center}
\end{figure}

In the proposed example the system has only one bound state. So,
with proper values of $M$ and $\beta$,
the diagonalization of $H$ results in one negative eigenvalue
$E_0$ and $M-1$ positive eigenvalues $E_j$ ($j=1,....,M-1$). The
corresponding wave functions
\begin{equation}
 \Psi_j=\sum_{m=0}^{M-1} a^j_m \phi_m  \hspace{0.5cm} j=1,....,M-1 \;\; ,
\end{equation}
are approximate solutions of $(H-E_j)\Psi_j=0$ in the interaction region.
As $r\rightarrow\infty$
they go to zero exponentially and therefore they do not represent a physical
scattering state.
The negative energy $E_0$ and the first three positive energy eigenvalues
($E_j$, $j=1,3$)
are shown in Fig.~\ref{fig:itfig1} as a function of $\beta$ in the case of $M=40$.
We observe the plateau already reached by $E_0$ for the values
of $\beta$ showed in the figure. We observe also the monotonic
behavior of the positive eigenvalues toward zero as $\beta$ decreases.
The corresponding eigenvectors
can be used to compute the integral relations of Eq.~\refeq{relint} and to
calculate the second order estimate of the phase-shifts
$\delta_j$ at the specific energies $E_j$.
This analysis is shown in Table~\ref{tab:ittab1} in which the
non linear parameter $\beta$ of the Laguerre basis has been fixed to $1.2$ fm$^{-1}$.
In the first row of the
table the ground state energy is given for different values of the
number $M$ of Laguerre polynomials. The stability of $E_0$ at the
level of $1$ keV is achieved already with $M=20$.
For a given value of $M$, $E_j$, with $j=1,2,3$,
are the first three positive eigenvalues.
The eigenvectors corresponding to positive energies approximate
the scattering states at the specific energies. Since the lowest
scattering state appears at zero energy, none of the positive eigenvalues
can reach this value for any finite values of $M$.
Defining
$ k^2_j=\frac{m}{\hbar^2}E_j$, the second order estimate for the phase
shift at each energy and at each value of $M$ is obtained as
\begin{eqnarray}
-&\frac{m}{\hbar^2} <\Psi_j|H-E|F_j>=B_j
 &{\rm with} \hspace{0.2cm} F_j=\sqrt{\frac{k_j}{4\pi}} \frac{\sin(k_jr)}{k_jr}  \cr
\cr
&\frac{m}{\hbar^2} <\Psi_j|H-E|\tilde G_j>=A_j
 &{\rm with} \hspace{0.2cm} \tilde G_j=f_{reg}\sqrt{\frac{k_j}{4\pi}}
\frac{\cos(k_jr)}{k_jr}  \cr
\cr
& [\tan\delta_j]^{2^{nd}} = B_j/A_j.
\label{rel2}
\end{eqnarray}

\begin{table}[h]
\caption{The two-nucleon bound state $E_0$ and the first three positive eigenvalues
$E_j$ $(j=1,3)$, as a function of the number of Laguerre polynomials $M$.
The second order estimates, $[\tan\delta_j]^{2^{nd}}$, obtained applying the integral
relations are given in each case and compared to exact results, $\tan\delta_j$.}
\begin{tabular}{lcccc}
\hline
 M    &  10       &  20       &  30      &  40       \cr
\hline
$E_0$ &-0.395079 &-0.397740  &-0.397743 &-0.397743  \cr
\hline
$E_1$ & 0.536349 & 0.116356  & 0.048091 & 0.026008  \cr
$[\tan\delta_1]^{2^{nd}}$
      &-1.507280 &-0.622242  &-0.392005 &-0.286479  \cr
$\tan\delta_1$
      &-1.522377 &-0.621938  &-0.392021 &-0.286480  \cr
\hline
$E_2$ & 1.984580 & 0.449655  & 0.190019 & 0.103503  \cr
$[\tan\delta_2]^{2^{nd}}$
      &-5.919685 &-1.353736  &-0.812313 &-0.584389  \cr
$\tan\delta_2$
      &-5.703495 &-1.354691  &-0.812270 &-0.584388  \cr
\hline
$E_3$ & 4.512635 & 0.994433  & 0.423117 & 0.231645  \cr
$[\tan\delta_3]^{2^{nd}}$
      &13.998124 &-2.451174  &-1.302799 &-0.908128  \cr
$\tan\delta_3$
      &12.684474 &-2.448343  &-1.302887 &-0.908131  \cr
\hline
\end{tabular}
\label{tab:ittab1}
\end{table}

On the other hand, as we are considering the $A=2$ system,
at each specified energy $E_j$ the phase shift $\tan\delta_j$
can be obtained
by solving the Schr\"odinger equation numerically. The two values,
$[\tan\delta_j]^{2^{nd}}$ and $\tan\delta_j$, are given in the Table~\ref{tab:ittab1}
at the corresponding energies as a function of $M$. We observe that, as
$M$ increases, the relative difference between the variational estimate
and the exact value reduces, for example at $M=40$ is around $10^{-6}$. In fact,
as $M$ increases, each eigenvector gives a better representation of
the exact wave function in the internal region and the second order
estimates, $[\tan\delta_j]^{2^{nd}}$ approach the exact result.

\begin{figure}
\begin{center}
\vspace{1.0cm}
\includegraphics[scale=0.35,angle=0]{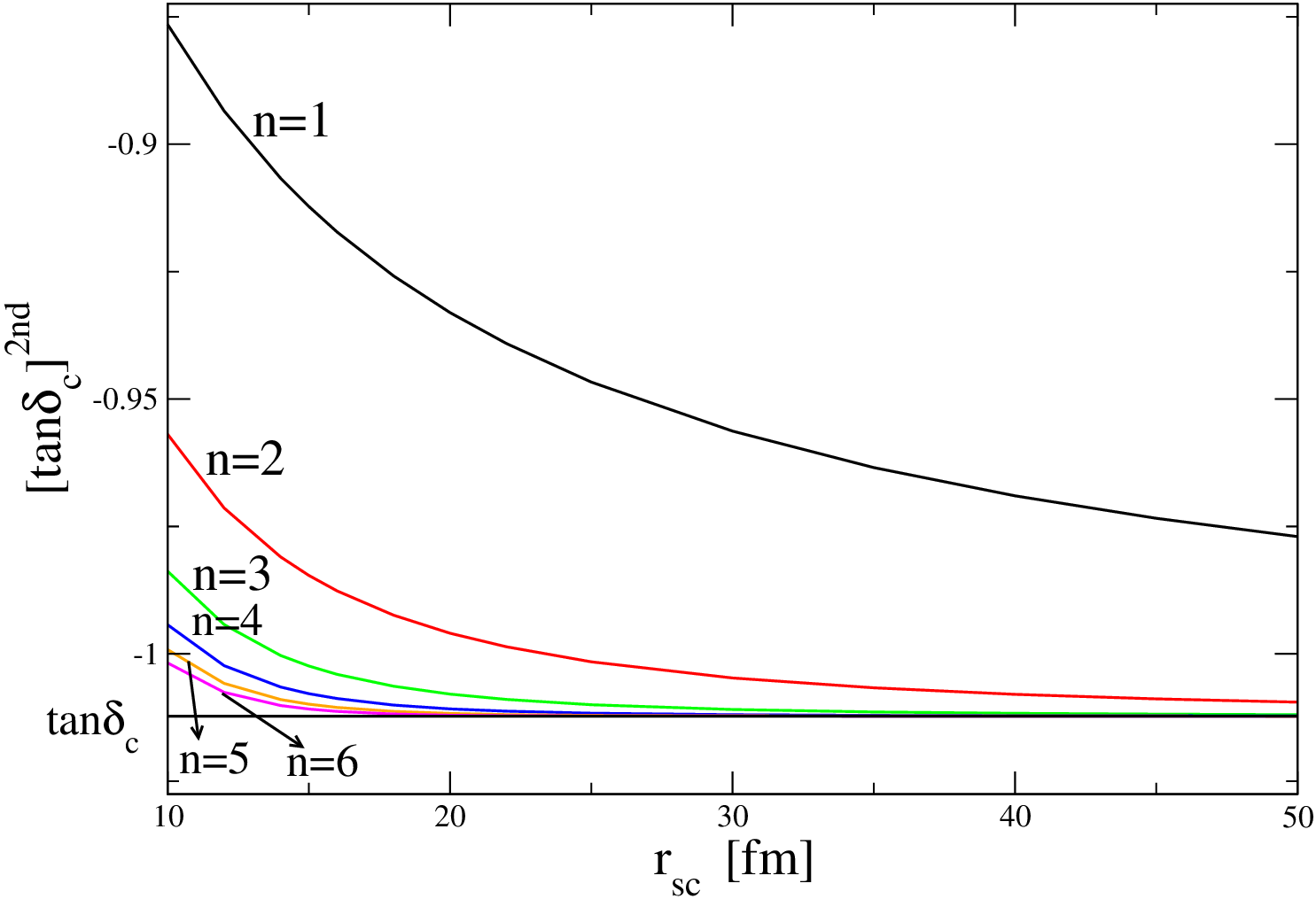}
\caption{The two-nucleon second order estimate $[\tan\delta_c]^{2^{nd}}$
as a function of $r_{sc}$ for different values of $n$. As a reference
the exact value for $\tan\delta_c$ is given as a straight line. }
\label{fig:itfig3}
\end{center}
\end{figure}

In a different application, the integral relations  can be used to calculate
the phase-shift of a process in which the two particles
interact through a short range potential plus the Coulomb potential,
imposing free asymptotic conditions to the wave function.
As an example we use the same two body potential used in the previous analysis
and add the Coulomb potential:
\begin{equation}
 V(r)=-V_0\exp{-(r/r_0)^2}+ \frac{e^2}{r}  \,\,\, .
\label{potc}
\end{equation}
For positive energies and $l=0$, the wave function behaves asymptotically as
\begin{equation}
 \Psi^{(c)}(r\rightarrow\infty)= AF_c(r)+BG_c(r)\;\; ,
\label{asympc}
\end{equation}
with $F_c(r),G_c(r)$ the regular and irregular Coulomb functions, respectively.
The phase-shift is $\tan\delta_c=B/A$. The KVP remains valid when the long range
Coulomb potential is considered and its form in terms of the integral
relations results:
\begin{eqnarray}
-&\frac{m}{\hbar^2} <\Psi^{(c)}_t|H-E|F_c>=B  \cr
\cr
&\frac{m}{\hbar^2} <\Psi^{(c)}_t|H-E|\tilde G_c>=A            \cr
\cr
& [\tan\delta_c]^{2^{nd}} = \frac{B}{A}\;\; .
\label{rel3}
\end{eqnarray}
with $\tilde G_c=f_{reg}G_c$ and $\Psi^{(c)}_t$ a trial wave function
behaving asymptotically as $\Psi^{(c)}$. Since 
$(H-E)|F_c>$ and $(H-E)|\tilde G_c>$ go to
zero outside the range of the short range potential, the integrals in
Eq.~\refeq{rel3} are negligible outside that region. Therefore, for
the computation of the phase-shift it is enough to require that 
$\Psi^{(c)}_t$ verifies
$(H-E)\Psi^{(c)}_t=0$, inside that region. To exploit this fact, we introduce the
following screened potential:
\begin{equation}
 V_{sc}(r)
=-V_0\exp{[-(r/r_0)^2]}+ \left[{\rm e}^{-(r/r_{sc})^n}\right]\frac{e^2}{r}\;\; .
\end{equation}
For specific values of $n$ and $r_{sc}$ it has the property of being
extremely close to the potential $V(r)$ of Eq.~\refeq{potc}
for $r<r_0$, with $r_0$ the range of the short
range potential. The screening factor ${\rm e}^{-(r/r_{sc})^n}$ cuts the
Coulomb potential for $r>r_{sc}$.
Using the potential $V_{sc}$ to describe a scattering process,
the wave function behaves asymptotically as
\begin{equation}
 \Psi_{n,r_{sc}}(r\rightarrow\infty)= AF(r)+BG(r)
\end{equation}
with $F,G$ from Eq.~\refeq{rel2},
since $V_{sc}$ is a short range potential. Solving the Schr\"odinger
equation for this potential,
it is possible to obtain the wave function $\Psi_{n,r_{sc}}$
for different values of $n$ and $r_{sc}$.
This wave function can be considered as a trial wave function for the problem
in which the Coulomb potential is unscreened. Accordingly it can be used as input
in Eq.~\refeq{rel3} to obtain a second order estimate
of the Coulomb phase-shift,
\begin{eqnarray}
-&\frac{m}{\hbar^2} <\Psi_{n,r_{sc}}|H-E|F_c>=B  \cr
\cr
&\frac{m}{\hbar^2} <\Psi_{n,r_{sc}}|H-E|\tilde G_c>=A            \cr
\cr
& [\tan\delta_c]^{2^{nd}} = \frac{B}{A}
\label{rel4}
\end{eqnarray}
where in $H$ the unscreened Coulomb potential is considered.
This estimate depends
on $n$ and $r_{sc}$ as the wave function does. In Fig.~\ref{fig:itfig3} the
second order estimate $[\tan\delta_c]^{2^{nd}}$ is shown as a function of
$r_{sc}$ for different values of $n$. The straight line is the exact value
of $\tan\delta_c$ obtained solving the Schr\"odinger equation.
We can observe that for $n\ge4$ and $r_{sc}>30$ fm
the second order estimate coincides with the exact results. In this example
the integral relations derived from the Kohn Variational Principle have
been used to extract a phase-shift in presence of the Coulomb potential
using wave functions with free asymptotic conditions.

Finally an application of the integral relations to the $A=3$ system is 
discussed. To this end we give the generalization of the integral
relations to the case in which more than one channel is open.
The coefficients $A$ and $B$ of Eq.~\refeq{rel2}
correspond to matrices
\begin{eqnarray}
B_{ij}&=&-\frac{m}{\hbar^2}<\Psi_i|H-E|F_j>  \cr
\cr
A_{ij}&=&\frac{m}{\hbar^2} <\Psi_i|H-E|{\widetilde G}_j> \cr
\cr
R^{2^{nd}}&=&A^{-1}B. \; \;\;\
\label{secondij}
\end{eqnarray}
with $R^{2^{nd}}$ the second order estimate of the scattering matrix
whose eigenvalues are the phase shifts and the indices $(i,j)$ indicate
the different asymptotic configurations accessible at the specific energy
under consideration. We consider $p-d$ scattering at $E_{lab}=3$ MeV
using the AV18 potential in the $J=1/2^+$ state. The corresponding
scattering matrix is a $2\times 2$ matrix. The corresponding phase-shift
and mixing parameters have been calculated using the PHH expansion
and are given in Table~\ref{tab:irtab2}. From the
previous discussion we have shown that it is possible to solve an
equivalent problem with a screened Coulomb potential, so with free
asymptotic conditions, and then use the integral relations to extract
the scattering matrix corresponding to the unscreened problem. This has
been done using Eq.~\refeq{secondij} and the results are given in
Table~\ref{tab:irtab2} using $r_{sc}=50$ fm and $n_{sc}=5$.
We observe a complete agreement between the two procedures.

\begin{table}[h]
\caption{Phase-shift and mixing parameters for $p-d$ scattering
at $E_{lab}=3$ MeV using the AV18 potential. Results using the
PHH expansion (second column) and using the integral relations
(last column)}
\begin{tabular}{lcc}
\hline
              & $p-d$         & Int.Rel.       \\
\hline
  $ ^4D_{1/2}$&$-3.563^\circ$ & $-3.562^\circ $ \\
  $ ^2S_{1/2}$&$-32.12^\circ$ & $-32.12^\circ $ \\
 $\eta_{1/2+}$&$1.100^\circ $ & $ 1.101^\circ $ \\
\hline
\end{tabular}
\label{tab:irtab2}
\end{table}

\section{Conclusions}
Stimulated by the fact that the commonly used TNF models
do not reproduce simultaneously
the triton and $^4$He binding energy and the $n-d$ doublet scattering length, we
have analyzed possible modifications of some of the TNF models usually used
in the description of light nuclei: the TM' and the URIX models. We have also
considered the recent N2LOL model. In each of these models we have 
varied the original parameters  so as to improve the
description of the mentioned quantities. Furthermore we have studied the 
description of some $p-d$ polarization observables at $E_{lab}=3$ MeV. We have
observed that the modification of the URIX produces a worse description of the
vector polarization observables due to the artificial increase of the strength
of the repulsive term. The analysis of the TM' model has put in evidence the
necessity of including a repulsive term. In the case of the N2LOL model a
fine tuning of the parameters was possible in order to have an acceptable
description of the triton and $^4$He binding energies and the $n-d$ doublet scattering 
length. Moreover, in the polarization observables we observe an improvement
in the vector analyzing powers and a slightly worse description of $T_{21}$. 
From this analysis we have established a connection between the short-range structure of
the TNF and the polarization observables at low energies.

In a different application, we have discussed the use of the integral
relations derived from the KVP in the description of scattering states.
Firstly we have shown the use of bound state like wave functions to compute
the scattering  matrix and, in the case of charged particles, the possibility 
of computing
phase-shifts using scattering wave functions with free asymptotic conditions,
obtained after screening the Coulomb interaction. Both problems are of
interest in the study of light nuclei.

\section{Acknowledgments}
This work has been done in collaboration with my colleagues in Pisa
M. Viviani, L. Girlanda and L.E. Marcucci, with C. Romero-Redondo
and E. Garrido (CSIC) and P. Barletta (UCL).

\section{Bibliography}

\label{biblio}

\end{document}